\pdfoutput=1

\documentclass[11pt]{article}

\usepackage[]{ACL2023}

\usepackage{times}
\usepackage{latexsym}

\usepackage[T1]{fontenc}

\usepackage[utf8]{inputenc}

\usepackage{microtype}

\usepackage{inconsolata}

\usepackage[utf8]{inputenc} 
\usepackage[T1]{fontenc}    
\usepackage{hyperref}       
\usepackage{amsfonts}       
\usepackage{graphicx}
\usepackage{amsthm}
\usepackage{amsmath}
\usepackage{subfigure}
\usepackage{multirow}
\usepackage{enumitem}
\usepackage{bbding}
\usepackage{colortbl}
\usepackage{arydshln}
\usepackage{booktabs}
\usepackage{tcolorbox}

\definecolor{pos}{RGB}{222, 242, 213}
\definecolor{neg}{RGB}{248, 202, 202}

\definecolor{lightgreen}{RGB}{224, 242, 213}
\definecolor{lightred}{RGB}{249,202,202}

\newcommand{\ra}[1]{\renewcommand{\arraystretch}{#1}} 
\newcommand{\paratitle}[1]{\vspace{0.8ex}\noindent\textbf{#1}}

\title{\includegraphics[width=1cm]{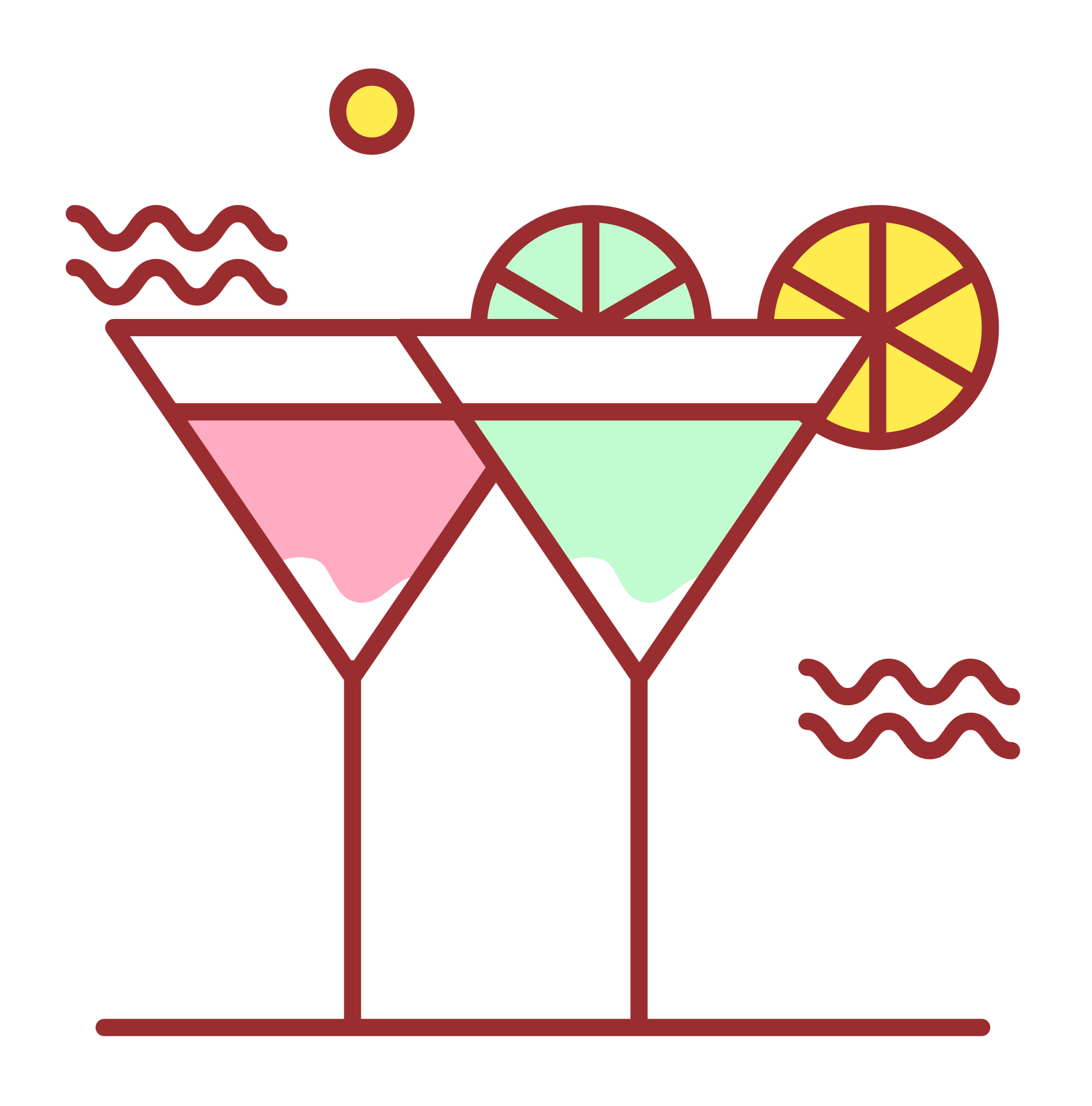} Cocktail: A Comprehensive Information Retrieval Benchmark \\ with LLM-Generated Documents Integration}

\author{
    Sunhao Dai\textsuperscript{1},
    Weihao Liu\textsuperscript{1},
    Yuqi Zhou\textsuperscript{1},
    Liang Pang\textsuperscript{2},
    Rongju Ruan\textsuperscript{3},
    Gang Wang\textsuperscript{3},\\
    \textbf{Zhenhua Dong\textsuperscript{3},
    Jun Xu\textsuperscript{1}\thanks{~~Corresponding author. Work partially done at Engineering Research Center of Next-Generation Intelligent Search and Recommendation, Ministry of Education.},
    Ji-Rong Wen\textsuperscript{1} } \\
    \textsuperscript{1}Gaoling School of Artificial Intelligence, Renmin University of China \\
    \textsuperscript{2}CAS Key Laboratory of AI Safety, Institute of Computing Technology, CAS \\
    \textsuperscript{3}Huawei Noah's Ark Lab\\
    \texttt{\{sunhaodai, junxu\}@ruc.edu.cn, } 
    \texttt{pangliang@ict.ac.cn}\\
}

\begin{document}
\maketitle
\begin{abstract}

The proliferation of Large Language Models (LLMs) has led to an influx of AI-generated content (AIGC) on the internet, transforming the corpus of Information Retrieval (IR) systems from solely human-written to a coexistence with LLM-generated content. The impact of this surge in AIGC on IR systems remains an open question, with the primary challenge being the lack of a dedicated benchmark for researchers. In this paper, we introduce Cocktail, a comprehensive benchmark tailored for evaluating IR models in this mixed-sourced data landscape of the LLM era. Cocktail consists of 16 diverse datasets with mixed human-written and LLM-generated corpora across various text retrieval tasks and domains. Additionally, to avoid the potential bias from previously included dataset information in LLMs, we also introduce an up-to-date dataset, named NQ-UTD, with queries derived from recent events. Through conducting over 1,000 experiments to assess state-of-the-art retrieval models against the benchmarked datasets in Cocktail, we uncover a clear trade-off between ranking performance and source bias in neural retrieval models, highlighting the necessity for a balanced approach in designing future IR systems. We hope Cocktail can serve as a foundational resource for IR research in the LLM era, with all data and code publicly available at \url{https://github.com/KID-22/Cocktail}.

\end{abstract}

\section{Introduction}

Information retrieval (IR) systems,  as the keystone in overcoming information overload, have seen widespread application across various domains, including search engines~\cite{li2014semantic}, question answering~\cite{karpukhin-etal-2020-dense}, dialog systems~\cite{chen2017survey}, etc.
A typical IR system aims at finding relevant documents or passages from a specific corpus in response to user's queries~\cite{li2022learning, zhao2022dense}.
Traditionally, IR benchmarks, notably MS MARCO~\cite{nguyen2016ms}, TREC~\cite{craswell2020overview}  and BEIR~\cite{thakur2021beir}, have exclusively utilized human-written content. 
However, the recent surge in Artificial Intelligence Generated Content (AIGC) facilitated by advanced Large Language Models (LLMs) has revolutionized the IR landscape~\cite{dai2024unifying}.
This evolution has broadened the scope of IR systems, which now encompass a hybrid corpus composed of both human-written and LLM-generated content~\cite{ai2023information, zhu2023large, dai2024unifying}, presenting new challenges and opportunities of IR in the LLM era.

For instance, recent studies~\cite{dai2023llms, xu2023ai} have shed light on a critical issue within neural retrieval models in the LLM era: a pronounced ``source bias''. 
This bias, characterized by the preferential ranking of LLM-generated content over semantically equivalent human-written content, poses a significant threat to the IR ecosystem. 
Hence, the need to comprehensively understand such impact of LLM-generated content across different IR models and diverse IR domains and tasks has become more pressing, especially with the escalating prevalence of LLM-generated content~\cite{hanley2023machine, bengio2023managing}. 
However, existing IR benchmarks either fail to reflect the real-world IR scenarios of the LLM era, as they solely contain human-written texts in their corpus, or provide limited datasets for exploring source bias. 
These shortcomings highlight the need for a comprehensive benchmark that accurately mirrors the current IR landscape, characterized by the integration of both human-written and LLM-generated texts within the corpus, to facilitate new research questions in this LLM era.

To fill this gap, we present a comprehensive benchmark tailored for IR in the LLM era, namely Cocktail, where the corpus contains both human-written and LLM-generated texts. 
Cocktail encompasses 16 retrieval datasets spanning different domains and tasks, enabling both in-domain and out-of-domain evaluation settings. 
To construct these datasets, we first select 15 widely used public human-written corpora from MS MARCO~\cite{nguyen2016ms}, TREC~\cite{craswell2020overview}, and BEIR~\cite{thakur2021beir}.
Then, based on these human-written corpora, we use Llama2~\cite{touvron2023llama} to rewrite each text to preserve semantic equivalence while introducing LLM-generated corpora.
Finally, we mix the original human-written corpora and the LLM-generated corpora to get the final Cocktail corpora and assign the same relevancy label for the corresponding query-document pairs. 
Furthermore, to address the potential biases introduced by the inherent knowledge of LLMs during the rewriting process, we collect an additional new dataset, NQ-UTD. This new dataset comprises 80 queries and 800 documents from recent events. It serves as a critical component of Cocktail, offering an essential perspective for assessing the performance of IR systems in the context of both pre- and post-LLM era datasets.

\begin{figure}[t]
\centering
\includegraphics[width=1\columnwidth]{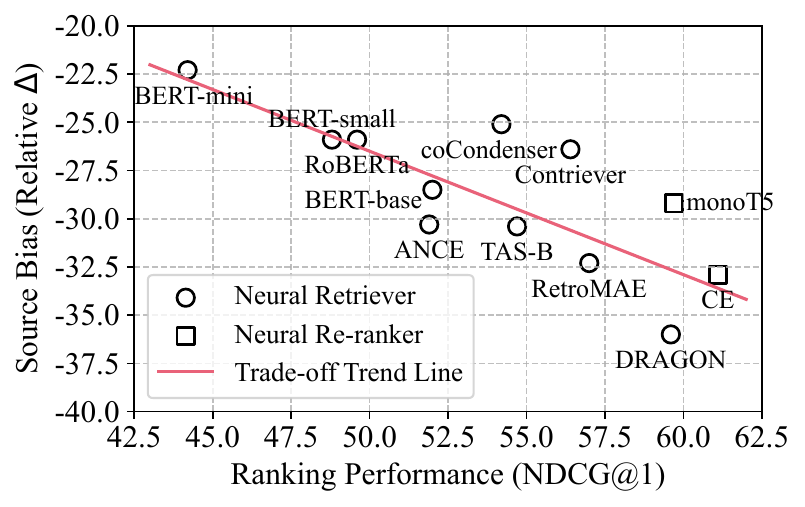}
\caption{Ranking performance versus source bias comparison on averaged results of 16 datasets benchmarked in Cocktail. A more negative Relative $\Delta$ signifies increased source bias towards LLM-generated content. The Pearson correlation coefficient between these two axes is $-0.772$ ($p$-value < 0.05), indicating a strong negative correlation. For brevity, we omit the `\%' symbol of the scores in all the tables and figures.}
\label{fig: intro}
\end{figure}

With the benchmarked diverse datasets in Cocktail, we then conduct comprehensive evaluations of over ten state-of-the-art (SOTA) retrieval models through more than 1,000 experiments. Our analysis firstly reinforces previous findings by \citet{dai2023llms}, highlighting a pervasive bias towards LLM-generated content across nearly all 16 datasets in Cocktail among all neural retrieval models. Furthermore, the results illustrated in \autoref{fig: intro} reveal a distinct trade-off between ranking performance and source bias within these SOTA neural models. 
This observation suggests that while striving for high performance, these models may rely on inherent shortcuts, failing to grasp true semantic relevance and resulting in severe source bias. Hence, future work requires better considering a balance between performance and bias mitigation in the design of next-generation IR models.

In summary, our contributions are as follows: 

(1) To the best of our knowledge, Cocktail is the first comprehensive benchmark with 16 datasets from a variety of domains and tasks tailed for IR research in the LLM era, where the corpus of each dataset contains both human-written texts and corresponding LLM-generated counterparts.

(2) We conduct extensive evaluations of state-of-the-art retrieval models using the Cocktail benchmark, assessing both retrieval accuracy and source bias. The evaluation tool, along with the codes, is open-sourced, facilitating ease of adaptation for evaluating new models and datasets.

(3) Extensive empirical studies reveal a clear trade-off between ranking performance and source bias in neural retrieval models. This finding underscores the importance of achieving a suitable balance between performance improvement and bias mitigation in future IR model designs.

\begin{table*}[t]
\centering
\aboverulesep=0pt
\belowrulesep=0pt
\resizebox{1\linewidth}{!}{
\renewcommand{\arraystretch}{1.3}
\begin{tabular}{c|c|c|c|c|c|rrr|rrr}
\toprule
\multirow{2}{*}{Dataset} & \multicolumn{3}{c|}{}         & Train    & Dev      & \multicolumn{3}{c|}{Test}    & \multicolumn{3}{c}{Avg. Word Length}        \\
\cmidrule(lr){2-4}  \cmidrule(lr){5-5} \cmidrule(lr){6-6}  \cmidrule(lr){7-9} \cmidrule(lr){10-12}
   & Domain   & Task   & Relevancy & \# Pairs & \# Query & \# Query & \# Corpus & Avg. D/Q & Query            & Human Doc & LLM Doc \\
\hline
\multicolumn{11}{c}{Collected Before the Emergence of LLM ($\sim$ - 2021/04)} \\
\hline
MS MARCO        & Misc.  & Passage-Retrieval     & Binary    & 532,663   & -        & 6,979     & 542,203    & 1.1      & 6.0              & 58.1      & 55.1    \\
DL19            & Misc.  & Passage-Retrieval     & Binary    & -        & -        & 43       & 542,203    & 95.4     & 5.4              & 58.1      & 55.1    \\
DL20            & Misc.  & Passage-Retrieval     & Binary    & -        & -        & 54       & 542,203    & 66.8     & 6.0              & 58.1      & 55.1    \\
TREC-COVID      & Bio-Medical & Bio-Medical IR & 3-level   & -        & -        & 50       & 128,585    & 430.1    & 10.6             & 197.6     & 165.9   \\
NFCorpus        & Bio-Medical & Bio-Medical IR & 3-level   & 110,575   & 324      & 323      & 3,633      & 38.2     & 3.3              & 221.0     & 206.7   \\
NQ              & Wikipedia   & Question Answering & Binary    & -        & -        & 3,446     & 104,194    & 1.2      & 9.2              & 86.9      & 81.0    \\
HotpotQA        & Wikipedia   & Question Answering & Binary    & 169,963   & 5447     & 7,405     & 111,107    & 2.0      & 17.7             & 67.9      & 66.6    \\
FiQA-2018       & Finance   & Question Answering  & Binary    & 14,045    & 499      & 648      & 57,450     & 2.6      & 10.8             & 133.2     & 107.8   \\
Touché-2020     & Misc.   & Argument Retrieval    & 3-level   & -        & -        & 49       & 101,922    & 18.4     & 6.6              & 165.4     & 134.4   \\
CQADupStack     & StackEx.  & Dup. Ques.-Retrieval  & Binary    & -        & -        & 1,563     & 39,962     & 2.4      & 8.5              & 77.2      & 72.0    \\
DBPedia         & Wikipedia  & Entity-Retrieval  & 3-level   & -        & 67       & 400      & 145,037    & 37.3     & 5.4              & 53.1      & 54.0    \\
SCIDOCS         & Scientific  & Citation-Prediction & Binary    & -        & -        & 1,000     & 25,259     & 4.7      & 9.4              & 169.7     & 161.8   \\
FEVER           & Wikipedia  & Fact Checking & Binary    & 140,079   & 6666     & 6,666     & 114,529    & 1.2      & 8.1              & 113.4     & 91.1    \\
Climate-FEVER   & Wikipedia  & Fact Checking & Binary    & -        & -        & 1,535     & 101,339    & 3.0      & 20.2             & 99.4      & 81.3    \\
SciFact         & Scientific  & Fact Checking & Binary    & 919      & -        & 300      & 5,183      & 1.1      & 12.4             & 201.8     & 192.7   \\
\hline
\multicolumn{11}{c}{Collected After the Emergence of LLM (2023/11 - 2024/01)} \\
\hline
NQ-UTD & Misc.    & Question Answering    & 3-level   & -        & -        & 80       & 800       & 3.7      & 12.1             & 101.1     & 94.7    \\ 
\bottomrule
\end{tabular}
}
\caption{Statistics of all 16  datasets in Cocktail benchmark. Avg. D/Q denotes the average number of relevant documents per query.}
\label{tab: dataset_stat}
\end{table*}

\section{Related Work}
\paratitle{IR meets Large Language Models.} Information retrieval (IR), the keystone of information access, has now significantly been reshaped by the emergence of large language models (LLMs)~\cite{zhao2023survey, ai2023information, dai2024unifying}. This intersection has manifested in two pivotal ways. 
On the one hand, much effort has been made to utilize the advanced capabilities of LLMs to refine the whole retrieval pipeline~\cite{zhu2023large}, including the integration of LLMs across various IR components, such as query rewriters~\cite{srinivasan2022quill}, retrievers~\cite{wang2023query2doc}, re-rankers~\cite{sun-etal-2023-chatgpt}, and readers~\cite{shi2023replug}.
On the other hand, the capacity of LLMs for generating human-like text at scale has shifted the landscape of searchable corpora, which now includes both human-written and LLM-generated texts. This evolution has introduced new challenges, most notably the emergence of source bias~\cite{dai2023llms, xu2023ai, dai2024unifying}, where neural retrievers exhibit a preference for LLM-generated content, potentially compromising the fairness and accuracy of search results. Moreover, the inclusion of LLM-generated content also raises concerns about privacy~\cite{yao2023survey} and the dissemination of misinformation~\cite{pan2023risk}. 
In this paper, we focus on the second line and aim to establish a comprehensive benchmark for evaluating IR models in the LLM era, which can help understand the impact of LLM-generated content on IR systems.

\paratitle{Related Benchmarks.} 
Historically, before the proliferation of LLM-generated content on the internet, several benchmarks were established for the evaluation of IR models, primarily utilizing corpora composed of human-written documents or passages. Notably, MS MARCOO~\cite{nguyen2016ms} and TREC~\cite{craswell2020overview} are widely used for supervised evaluation in IR research. Similarly, BEIR~\cite{thakur2021beir} presents a diverse benchmark incorporating 18 datasets from various IR domains and tasks, tailored to zero-shot evaluation. Despite their contributions to advancing IR systems, these benchmarks fall short of reflecting the real-world scenarios of the current LLM era due to the absence of LLM-generated content in their corpora~\cite{dai2023llms, xu2023ai, dai2024unifying}. This gap underscores the necessity for new benchmarks that include both human-written and LLM-generated texts, offering a more comprehensive and realistic evaluation environment to navigate the challenges and opportunities presented by the integration of LLMs into IR.

\begin{figure*}[t]
\centering
\includegraphics[width=1\textwidth]{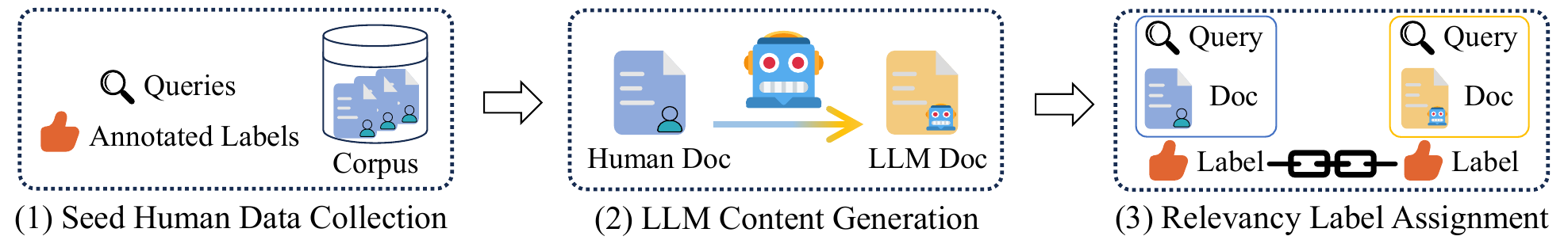}
\caption{An overview of the dataset construction pipeline involved in Cocktail.}
\label{fig: framework}
\end{figure*}

\section{Benchmarking Retrieval Datasets}

Cocktail establishes a comprehensive benchmark tailored for the evaluation of IR models in the LLM era, characterized by corpora containing both human-written and LLM-generated content. This benchmark aims to access the performance and biases of existing retrieval models while encouraging the creation of future IR systems that excel in robustness and generalization across diverse scenarios in the LLM era. To achieve this, we collect and construct 16 IR datasets, incorporating 15 datasets from the pre-LLM era and one newly developed dataset to ensure a wide representation of domains and tasks. 
The statistics for the 16 datasets benchmarked in Cocktail are summarized in \autoref{tab: dataset_stat}. We also list the dataset website links and the corresponding licenses in Appendix \autoref{tab:dataset_licenses}.

\subsection{Dataset Construction}
A key concern in constructing datasets with LLMs is the potential for LLMs to have prior knowledge about queries, which could lead to an unfair evaluation. To address this, following previous works~\cite{dai2023llms}, we choose rewriting documents without incorporating query-related information rather than a full generation from LLMs with given queries, ensuring that any detected source bias is a genuine reflection of model preferences. Moreover, by avoiding full document generation based on queries, this approach also simplifies controlling the text length across different sources, ensuring consistency and further reducing potential biases in the evaluation.
Specifically, we construct each dataset in Cocktail following a three-step process: 1) collecting seed datasets with human-written corpus and alongside relevancy labels for given queries; 2) leveraging LLMs to generate corresponding LLM-generated corpus with the human-written corpus as inputs; and 3) assigning relevancy labels to the LLM-generated content and given queries, ensuring seamless integration into the benchmark. An overview of our construction pipeline is shown in \autoref{fig: framework}.

\paratitle{Seed Human Datasets.} 
To cover as diverse IR domains and tasks as possible, we select the widely-used 15 datasets from MS MARCO~\cite{nguyen2016ms}, TREC~\cite{craswell2020overview}, and BEIR~\cite{thakur2021beir} benchmark as our human-written corpus. These datasets span six domains and eight distinct tasks, facilitating comprehensive evaluations under both in-domain and out-of-domain settings. Moreover, as suggested in \citet{sun-etal-2023-chatgpt}, we also construct a new test dataset NQ-UTD to ensure that relevance annotations have not been learned by LLMs. 
NQ-UTD comprises 80 queries distributed across eight domains, sourcing from recent events (from Nov. 2023 to Jan. 2024). We verify that neither GPT-4 nor Gemini, two of the most advanced LLMs to date, can not answer the questions within NQ-UTD, validating no prior knowledge of these questions in LLM.
For details about the collection and annotation processes of NQ-UTD, along with detailed descriptions of all 16 datasets, please refer to Appendix~\ref{app: datasets_details}.

\paratitle{LLM-Generated Corpus.}
To ensure a fair evaluation of the impact of incorporating LLM-generation content, we utilize the widely used \texttt{llama-2-7b-chat} to rewrite each human-written text without changing its semantic information, with the following instructions: ``\textit{Original Text: \{\{human-written text\}\} Please rewrite the above given text. Your answer must be formatted as follows: Rewritten Text: <your rewritten text>.}''  This post-instruction strengthens the instruction-following capabilities of LLMs, proving particularly effective for processing and generating responses to long input texts~\cite{liu2023instruction}.

\paratitle{Relevancy Label Assignment.}
Upon the creation of the LLM-generated corpus for each dataset, we assign the relevancy labels from the original <query, human-written doc> pairs to the new <query, LLM-generated doc> pairs. This process is underpinned by the premise that both sources of the content preserve nearly identical semantic information, which will be further verified in the following section.

All the datasets in Cocktail are organized in a standard format (corpus, queries, qrels) akin to the BEIR benchmark~\cite{thakur2021beir}, facilitating ease of use and comparison. Examples for each dataset are showcased in Appendix \autoref{tab: dataset_examples}. More details about the data processing and quality control are provided in Appendix~\ref{app: data_process}.

\subsection{Dataset Statistics and Analysis}
As shown in \autoref{tab: dataset_stat}, there are minimal differences in average word length between human-written and LLM-generated documents, with the latter being marginally shorter. 
The detailed text length distribution, depicted in Appendix \autoref{fig: stats_length}, further confirms the negligible variance in text length. 
Additionally, term-based Jaccard similarity and overlap distributions between LLM-generated and human-written documents, visualized in Appendix \autoref{fig: stats_jaccard_and_overlap}, show a noticeable distinction in terms despite similar semantic content.

Following practices in \citet{dai2023llms}, our dataset construction process involved feeding only the original human-written text to the LLM for rewriting, without inputting any query-specific information to avoid introducing additional query-related bias. 
Furthermore, we employed the OpenAI embedding model\footnote{{text-embedding-ada-002}: \url{https://platform.openai.com/docs/guides/embeddings}} to obtain semantic embeddings for both text sources, comparing them through cosine similarity. 
For calibrating the results, we also compare the semantic similarity between randomly selected <human-written doc, LLM-generated doc> pairs and matching pairs <human-written doc, LLM-generated doc>.
These comparisons, presented in Appendix \autoref{fig: stats_cosine_similarity} and \autoref{tab: calibration}, demonstrate a high similarity between the LLM-generated document and the corresponding original human-written counterpart, indicating successful semantic retention of the original human-written text in the LLM-generated texts. Additionally, evaluations of several retrieval models on sole human-written or LLM-generated corpora showed consistent performance, as seen in Appendix \autoref{tab: main_sole_ndcg_1}. 
Moreover, we also conduct human evaluations for quality verification for each dataset in Appendix \autoref{tab: human_eval_quality}.
These observations reinforce confidence in the quality of our newly constructed datasets, suggesting that the LLM-generated content maintains semantics comparable to human-written texts for IR tasks.
The detailed dataset analysis is provided in Appendix~\ref{app: data_ana_statis}.

\begin{table*}[t]
\centering
\resizebox{1.0\textwidth}{!}{
\renewcommand{\arraystretch}{1.25}
\begin{tabular}{c|c|cccccccc|cc|cc}
\hline\hline
\multirow{2}{*}{Model ($\rightarrow$)} &  Lexical & \multicolumn{8}{c|}{Neural Retrievers}  & \multicolumn{2}{c|}{Neural Re-rankers} \\

     & BM25  & BERT  & RoBERTa & ANCE    & TAS-B      & Contriever & coCondenser & RetroMAE & DRAGON & CE     & monoT5  &  \\
\hline
PLM           & -     & BERT  & RoBERTa & RoBERTa & DistilBERT & BERT       &    BERT     &     BERT      &   BERT      & MiniLM & T5     &  \multicolumn{2}{c}{Average}\\
\# Paras      &  -      & 110M  & 125M    & 125M    &    66M        & 110M       &   110M   &    110M      &   110M       &    66M    &  220M      &    All     & Neural    \\
\hline
\multicolumn{14}{c}{Supervised Evaluation (In-Domain Datasets Collected in Pre-LLM Era)} \\
\hline
MS MARCO   & 38.8 & 51.6 & 51.9   & 52.8   & 54.4 & 55.6      & 55.3       & 55.7    & \textbf{59.1}  & \underline{57.8} & 55.3 & 53.5 & 54.9   \\
DL19       & 57.8 & \underline{78.3} & 76.4   & 72.5   & 71.7 & 72.1      & 76.4       & 77.5    & \underline{78.3}  & \textbf{81.0} & 74.0 & 74.2 & 75.8   \\
DL20       & 55.3 & \underline{79.9} & 76.9   & 73.2   & 72.5 & 73.8      & 79.3       & 76.5    & \textbf{82.4}  & 73.5 & 71.0 & 74.0 & 75.9  \\

\hline
\multicolumn{14}{c}{Zero-shot Evaluation (Out-of-Domain Datasets Collected in Pre-LLM Era)} \\
\hline
TREC-COVID    & 67.0 & 67.0 & 62.0   & 68.0   & 65.0 & 64.0      & 70.0       & 72.0    & 75.0  & \underline{77.0} & \textbf{82.0} & 69.9 & 70.2   \\
NFCorpus      & \underline{45.4} & 39.2 & 36.4   & 35.3   & 41.5 & 42.6      & 43.5       & 41.6    & 42.7  & \textbf{50.2} & 44.1 & 42.0 & 41.7  \\
NQ            & 45.7 & 62.5 & 60.0   & 60.0   & 65.0 & 69.6      & 65.6       & 67.5    & \underline{70.4}  & 69.1 & \textbf{71.3} & 64.2 & 66.1   \\
HotpotQA   & 84.3 & 75.3 & 62.9   & 72.1   & 84.6 & 88.1      & 82.1       & 88.5    & 89.1  & \textbf{93.9} & \underline{89.8} & 82.8 & 82.6   \\
FiQA-2018     & 24.4 & 24.7 & 24.9   & 28.9   & 28.1 & 34.0      & 27.2       & 31.8    & \underline{36.3}  & 35.5 & \textbf{40.9} & 30.6 & 31.2   \\
Touché-2020   & \textbf{57.1} & 39.8 & 42.9   & 44.9   & 44.9 & 40.8      & 36.7       & 44.9    & 52.0  & \underline{56.1} & 55.1 & 46.8 & 45.8   \\
CQADupStack   & 28.6 & 26.6 & 26.9   & 31.3   & 23.4 & 34.2      & 33.2       & 31.3    & \underline{36.0}  & 34.7 & \textbf{36.8} & 31.2 & 31.4   \\
DBPedia       & 36.6 & 55.3 & 50.6   & 48.0   & 56.3 & 62.1      & 56.9       & 56.8    & \textbf{59.6}  & \underline{58.6} & 49.3 & 53.6 & 55.3   \\
SCIDOCS    & 16.2 & 12.3 & 11.2   & 14.5   & 16.0 & 16.6      & 14.5       & 16.2    & 17.9  & \underline{19.0} & \textbf{19.5} & 15.8 & 15.8  \\
FEVER         & 65.6 & 80.6 & 75.9   & 80.4   & 83.1 & 86.4      & 75.8       & 87.6    & \underline{88.0}  & \textbf{90.2} & 76.3 & 80.9 & 82.4   \\
Climate-FEVER & 25.2 & 26.4 & 27.4   & 28.3   & 32.7 & 31.4      & 28.1       & 32.3    & \underline{34.0}  & \textbf{35.2} & 33.8 & 30.4 & 31.0  \\
SciFact       & 55.7 & 38.0 & 40.3   & 40.7   & 53.7 & 55.0      & 50.0       & 52.3    & 55.3  & \underline{56.3} & \textbf{65.3} & 51.2 & 50.7  \\
\hline
\multicolumn{14}{c}{Zero-shot Evaluation (Out-of-Domain Datasets Collected in the LLM Era)} \\
\hline
NQ-UTD         & 76.9 & 74.4 & 66.9   & 78.8   & 81.9 & 75.6      & 73.1       & 78.8    & 76.9  & \underline{89.4} & \textbf{90.0} & 78.4 & 78.6   \\
\hline
\multicolumn{14}{c}{Averaged Result} \\
\hline
Supervised & 50.6 & 69.9 & 68.4 & 66.2 & 66.2 & 67.2 & 70.3 & 69.9 & \textbf{73.3} & \underline{70.8} & 66.8 & 67.2 & 68.9  \\
Zero-shot  &  48.4 & 47.9 & 45.3 & 48.6 & 52.0 & 53.9 & 50.5 & 54.0 & 56.4 & \textbf{58.9} & \underline{58.0} & 52.1 & 52.5  \\
All   & 48.8 & 52.0 & 49.6   & 51.9   & 54.7 & 56.4      & 54.2       & 57.0    & 59.6  & \textbf{61.1} & \underline{59.7} & 55.0 & 55.6 \\
\hline\hline
\end{tabular}
}
\caption{Overall ranking performance (NDCG@1) across all benchmarked datasets in Cocktail. The second-to-last column is the average result across all models, while the last column is the average for all neural retrieval models. The \textbf{best performed} result for each dataset is marked in bold, and the \underline{second best} is underlined. }
\label{tab: main_mix_ndcg_1}
\end{table*}

\begin{table*}[t]
\centering
\resizebox{1.0\textwidth}{!}{
\renewcommand{\arraystretch}{1.25}
\begin{tabular}{c|c|cccccccc|cc|cc}
\hline\hline
\multirow{2}{*}{Model ($\rightarrow$)} &  Lexical & \multicolumn{8}{c|}{Neural Retrievers}  & \multicolumn{2}{c|}{Neural Re-rankers} \\

     & BM25  & BERT  & RoBERTa & ANCE    & TAS-B      & Contriever & coCondenser & RetroMAE & DRAGON & CE     & monoT5  &  \\
\hline
PLM           & -     & BERT  & RoBERTa & RoBERTa & DistilBERT & BERT       &    BERT     &     BERT      &   BERT      & MiniLM & T5     &  \multicolumn{2}{c}{Average}\\
\# Paras      &  -      & 110M  & 125M    & 125M    &    66M        & 110M       &   110M   &    110M      &   110M       &    66M    &  220M      &    All     & Neural    \\
\hline
\multicolumn{14}{c}{Supervised Evaluation (In-Domain Datasets Collected in Pre-LLM Era)} \\
\hline
MS MARCO   & \cellcolor{pos}72.2  & \cellcolor{neg}-13.2 & \cellcolor{neg}-18.9   & \cellcolor{pos}1.1     & \cellcolor{neg}-29.0      & \cellcolor{neg}-10.8      & \cellcolor{pos}5.1         & \cellcolor{neg}-20.5    & \cellcolor{neg}-18.2  & \cellcolor{neg}-26.3  & \cellcolor{neg}-7.8   & \cellcolor{neg}-6.0    & \cellcolor{neg}-13.8   \\
DL19       & \cellcolor{pos}108.7 & \cellcolor{neg}-51.3 & \cellcolor{pos}11.0    & \cellcolor{neg}-0.8    & \cellcolor{neg}-55.0      & \cellcolor{neg}-21.4      & \cellcolor{neg}-39.6       & \cellcolor{neg}-53.9    & \cellcolor{neg}-111.1 & \cellcolor{neg}-18.3  & \cellcolor{neg}-53.7  & \cellcolor{neg}-25.9   & \cellcolor{neg}-39.4   \\
DL20       & \cellcolor{pos}101.6 & \cellcolor{neg}-76.5 & \cellcolor{neg}-63.5   & \cellcolor{neg}-2.5    & \cellcolor{neg}-11.0      & \cellcolor{neg}-10.8      & \cellcolor{neg}-38.1       & \cellcolor{neg}-43.7    & \cellcolor{neg}-59.2  & \cellcolor{neg}-25.3  & \cellcolor{neg}-11.3  & \cellcolor{neg}-21.8   & \cellcolor{neg}-34.2   \\
\hline
\multicolumn{14}{c}{Zero-shot Evaluation (Out-of-Domain Datasets Collected in Pre-LLM Era)} \\
\hline
TREC-COVID    & \cellcolor{pos}32.8  & \cellcolor{neg}-62.7 & \cellcolor{neg}-64.5   & \cellcolor{neg}-58.8   & \cellcolor{neg}-95.4      & \cellcolor{neg}-87.5      & \cellcolor{neg}-68.6       & \cellcolor{neg}-66.7    & \cellcolor{neg}-45.3  & \cellcolor{neg}-64.9  & \cellcolor{neg}-63.5  & \cellcolor{neg}-58.6   & \cellcolor{neg}-67.8   \\
NFCorpus      & \cellcolor{neg}-29.5 & \cellcolor{neg}-30.6 & \cellcolor{neg}-50.5   & \cellcolor{neg}-23.2   & \cellcolor{neg}-44.8      & \cellcolor{neg}-99.5      & \cellcolor{neg}-49.2       & \cellcolor{neg}-17.3    & \cellcolor{neg}-37.9  & \cellcolor{neg}-66.1  & \cellcolor{neg}-38.9  & \cellcolor{neg}-44.3   & \cellcolor{neg}-45.8   \\
NQ            & \cellcolor{neg}-17.9 & \cellcolor{neg}-26.6 & \cellcolor{neg}-16.7   & \cellcolor{neg}-12.7   & \cellcolor{neg}-41.8      & \cellcolor{neg}-37.9      & \cellcolor{neg}-25.0       & \cellcolor{neg}-25.2    & \cellcolor{neg}-47.2  & \cellcolor{neg}-61.6  & \cellcolor{neg}-33.6  & \cellcolor{neg}-31.5   & \cellcolor{neg}-32.8   \\
HotpotQA   & \cellcolor{pos}51.0  & \cellcolor{neg}-1.1  & \cellcolor{neg}-3.5    & \cellcolor{neg}-13.6   & \cellcolor{pos}0.2        & \cellcolor{neg}-5.7       & \cellcolor{neg}-5.4        & \cellcolor{pos}5.2      & \cellcolor{neg}-8.5   & \cellcolor{pos}36.6   & \cellcolor{pos}14.8   & \cellcolor{pos}6.4     & \cellcolor{pos}1.9     \\
FiQA-2018     & \cellcolor{neg}-8.2  & \cellcolor{neg}-38.1 & \cellcolor{pos}8.8     & \cellcolor{neg}-33.3   & \cellcolor{neg}-6.4       & \cellcolor{neg}-38.3      & \cellcolor{neg}-29.4       & \cellcolor{neg}-52.8    & \cellcolor{neg}-12.7  & \cellcolor{neg}-42.3  & \cellcolor{neg}-35.7  & \cellcolor{neg}-26.2   & \cellcolor{neg}-28.0   \\
Touché-2020   & \cellcolor{neg}-21.4 & \cellcolor{neg}-25.6 & \cellcolor{neg}-76.0   & \cellcolor{neg}-36.1   & \cellcolor{neg}-36.1      & \cellcolor{neg}-29.8      & \cellcolor{pos}10.9        & \cellcolor{neg}-9.4     & \cellcolor{neg}-66.4  & \cellcolor{neg}-127.3 & \cellcolor{neg}-66.4  & \cellcolor{neg}-44.0   & \cellcolor{neg}-46.2   \\
CQADupStack   & \cellcolor{pos}22.4  & \cellcolor{neg}-45.1 & \cellcolor{neg}-39.4   & \cellcolor{neg}-19.8   & \cellcolor{neg}-10.3      & \cellcolor{neg}-22.2      & \cellcolor{neg}-6.6        & \cellcolor{neg}-67.1    & \cellcolor{neg}-24.4  & \cellcolor{neg}-30.5  & \cellcolor{neg}-8.7   & \cellcolor{neg}-22.9   & \cellcolor{neg}-27.4   \\
DBPedia       & \cellcolor{pos}18.6  & \cellcolor{neg}-5.4  & \cellcolor{neg}-19.3   & \cellcolor{neg}-25.8   & \cellcolor{pos}2.5        & \cellcolor{pos}4.5        & \cellcolor{neg}-11.6       & \cellcolor{neg}-24.6    & \cellcolor{neg}-21.5  & \cellcolor{neg}-13.3  & \cellcolor{pos}5.5    & \cellcolor{neg}-8.2    & \cellcolor{neg}-10.9   \\
SCIDOCS    & \cellcolor{pos}2.5   & \cellcolor{pos}21.1  & \cellcolor{neg}-21.4   & \cellcolor{neg}-9.7    & \cellcolor{neg}-20.0      & \cellcolor{neg}0.0        & \cellcolor{neg}-26.2       & \cellcolor{neg}-22.2    & \cellcolor{neg}-16.8  & \cellcolor{neg}-27.4  & \cellcolor{neg}-35.9  & \cellcolor{neg}-14.2   & \cellcolor{neg}-15.8   \\
FEVER         & \cellcolor{neg}-26.2 & \cellcolor{pos}2.5   & \cellcolor{neg}-2.4    & \cellcolor{neg}-87.1   & \cellcolor{neg}-16.8      & \cellcolor{neg}-20.4      & \cellcolor{neg}-22.2       & \cellcolor{neg}-1.4     & \cellcolor{neg}-26.8  & \cellcolor{neg}-4.4   & \cellcolor{pos}11.5   & \cellcolor{neg}-17.6   & \cellcolor{neg}-16.7   \\
Climate-FEVER & \cellcolor{pos}6.3   & \cellcolor{neg}-15.2 & \cellcolor{neg}-16.1   & \cellcolor{neg}-109.9  & \cellcolor{neg}-22.6      & \cellcolor{neg}-12.7      & \cellcolor{neg}-17.8       & \cellcolor{neg}-15.5    & \cellcolor{neg}-10.6  & \cellcolor{pos}5.1    & \cellcolor{neg}-89.0  & \cellcolor{neg}-27.1   & \cellcolor{neg}-30.4   \\
SciFact       & \cellcolor{pos}1.1   & \cellcolor{neg}-52.6 & \cellcolor{neg}-14.9   & \cellcolor{neg}-29.6   & \cellcolor{neg}-53.3      & \cellcolor{pos}1.5        & \cellcolor{neg}-21.6       & \cellcolor{neg}-29.4    & \cellcolor{neg}-38.7  & \cellcolor{neg}-8.2   & \cellcolor{neg}-38.2  & \cellcolor{neg}-25.8   & \cellcolor{neg}-28.5   \\
\hline
\multicolumn{14}{c}{Zero-shot Evaluation (Out-of-Domain Datasets Collected in the LLM Era)} \\
\hline
NQ-UTD         & \cellcolor{pos}37.2  & \cellcolor{neg}-35.5 & \cellcolor{neg}-27.8   & \cellcolor{neg}-22.3   & \cellcolor{neg}-47.1      & \cellcolor{neg}-31.4      & \cellcolor{neg}-56.3       & \cellcolor{neg}-73.1    & \cellcolor{neg}-30.9  & \cellcolor{neg}-51.9  & \cellcolor{neg}-16.9  & \cellcolor{neg}-32.4   & \cellcolor{neg}-39.3   \\
\hline
\multicolumn{14}{c}{Averaged Result} \\
\hline
Supervised    & \cellcolor{pos}94.2  & \cellcolor{neg}-47.0 & \cellcolor{neg}-23.8   & \cellcolor{neg}-0.7    & \cellcolor{neg}-31.7      & \cellcolor{neg}-14.3      & \cellcolor{neg}-24.2       & \cellcolor{neg}-39.4    & \cellcolor{neg}-62.8  & \cellcolor{neg}-23.3  & \cellcolor{neg}-24.3  & \cellcolor{neg}-17.9   & \cellcolor{neg}-29.2   \\
Zero-shot     & \cellcolor{pos}5.3   & \cellcolor{neg}-24.2 & \cellcolor{neg}-26.4   & \cellcolor{neg}-37.1   & \cellcolor{neg}-30.1      & \cellcolor{neg}-29.2      & \cellcolor{neg}-25.3       & \cellcolor{neg}-30.7    & \cellcolor{neg}-29.8  & \cellcolor{neg}-35.1  & \cellcolor{neg}-30.4  & \cellcolor{neg}-26.6   & \cellcolor{neg}-29.8   \\
All    &  \cellcolor{pos}22.0  & \cellcolor{neg}-28.5 & \cellcolor{neg}-25.9   & \cellcolor{neg}-30.3   & \cellcolor{neg}-30.4      & \cellcolor{neg}-26.4      & \cellcolor{neg}-25.1       & \cellcolor{neg}-32.3    & \cellcolor{neg}-36.0  & \cellcolor{neg}-32.9  & \cellcolor{neg}-29.2  & \cellcolor{neg}-25.0   & \cellcolor{neg}-29.7  
 \\
\hline\hline
\end{tabular}
}
\caption{Overall source bias evaluation w.r.t. $\text{Relative}~\Delta$ (NDCG@1) across all benchmarked datasets in Cocktail. The \colorbox{pos}{numbers} (i.e., $\text{Relative}~~\Delta > 0$) suggest that retrieval models generally prefer human-written content while the \colorbox{neg}{numbers} (i.e., $\text{Relative}~~\Delta \leq 0$) indicate retrieval models prefer LLM-generated content.}
\label{tab: main_rela_ndcg_1}
\end{table*}

\section{Benchmarking Evaluation Protocol}
\paratitle{Evaluation Framework.} To standardize the assessment of IR models within the LLM era, we also develop a Python framework that combines user-friendliness with comprehensive evaluation capabilities. Built on the foundation of the BEIR~\cite{thakur2021beir}, our framework inherits its best features, including the ability to easily replicate experiments from open-sourced repositories and incorporate new models and datasets. A key innovation of our framework is its ability to conduct evaluations using either individual or mixed corpora, accommodating the mixture of human-written and LLM-generated content characteristic of the LLM era. These features make our framework an invaluable tool for advancing IR research and application in both academia and industry. 
A detailed description and all the scripts of our evaluation framework are provided in the link \url{https://github.com/KID-22/Cocktail}.

\paratitle{Evaluation Metrics.}
Aligned with the BEIR benchmark~\cite{thakur2021beir}, we select Normalized Discounted Cumulative Gain (NDCG@$K$) as our primary metric to assess retrieval accuracy, given its robustness in capturing the effectiveness of IR systems across tasks with binary and graded relevance judgments.
Following previous studies~\cite{dai2023llms, xu2023ai}, we choose $K=1$ since the top-$1$ item in the retrieved list is most likely to be viewed and clicked by users. To provide a more comprehensive evaluation, we also report results on $K=3$ and $K=5$ in Appendix~\ref{app: more_results}.

Moreover, in the multi-sourced corpus shaped by LLMs,  the evaluation of IR systems necessitates not only mere accuracy metrics but also a critical assessment of source bias~\cite{dai2023llms}. This bias, as evidenced by ranking LLM-generated content higher than human-written content, poses a significant challenge in today’s IR ecosystem. To quantify and normalize source bias in different datasets, we follow previous works~\cite{dai2023llms,xu2023ai} and adopt the $\text{Relative}~~\Delta$. This metric captures the relative percentage difference in NDCG scores between human-written and LLM-generated content, which is defined as:
\begin{equation*}
\scalebox{1}{
$
\text{Relative $\Delta$} = \frac{\texttt{NDCG}_\text{Human} - \texttt{NDCG}_\text{LLM}}{(\texttt{NDCG}_\text{Human} +\texttt{NDCG}_\text{LLM}) / 2} \times 100\%,  
$}
\end{equation*}
where the $\texttt{NDCG}_\text{Human}$ and $\texttt{NDCG}_\text{LLM}$ denote the NDCG scores attributed to human-written and LLM-generated content, respectively. Note that $\text{Relative}~\Delta > 0 $ indicates that IR models rank human-written content higher than LLM-generated content, and conversely, $\text{Relative}~\Delta < 0 $ indicates the opposite tendency. The absolute value of $\text{Relative}~\Delta$ reflects the extent of source bias.

\section{Benchmarking Retrieval Models}
In this section, we delve into the evaluation and analysis of various retrieval models utilizing the constructed Cocktail benchmarked datasets.

\subsection{Retrieval Models}
Following the BEIR benchmark~\cite{thakur2021beir}, we focus on evaluating the advanced state-of-the-art transformer-based neural retrieval models. Besides the widely used \textbf{lexical retriever} {BM25}~\cite{robertson2009probabilistic}, our experiments include two main types of neural retrieval models:

\paratitle{Neural Retriever.} We utilize and fine-tune the following two most commonly used pre-trained language models on the MS MARCO dataset~\cite{nguyen2016ms} using the official training script\footnote{\url{https://github.com/beir-cellar/beir/blob/main/examples/retrieval/training/train_msmarco_v3.py}} from BEIR: 1) BERT~\cite{devlin-etal-2019-bert}; 2) RoBERTa~\cite{liu2019roberta}. Additionally, we evaluate the performance of state-of-the-art models also trained on MS MARCO, employing officially released checkpoints:
3) {ANCE}~\cite{xiong2020approximate}; 4) {TAS-B}~\cite{hofstatter2021efficiently}; 5) {Contriever}~\cite{izacard2022unsupervised}; 6) {coCondenser}~\cite{gao2022unsupervised}; 7) {RetroMAE}~\cite{xiao2022retromae} ; 8) {DRAGON}~\cite{lin2023train}.

\paratitle{Neural Re-ranker.} For re-ranking, we employ two state-of-the-art models with their publicly available official pre-trained checkpoints: 1) CE~\cite{wang2020minilm}; 2) monoT5~\cite{raffel2020exploring}.

In our experiments, unless specified otherwise, neural re-rankers re-rank the top-100 documents retrieved by BM25.
Detailed information on the benchmarked models, including the publicly available official pre-trained checkpoints and implementation details, can be found in Appendix~\ref{app: model_details}.

\subsection{Benchmarked Results} \label{sec: main_res}

We conduct extensive evaluations with more than 1,000 experiments on the Cocktail benchmarked datasets. 
The results of retrieval accuracy and source bias across all benchmarked retrieval models and datasets are reported in \autoref{tab: main_mix_ndcg_1} and \autoref{tab: main_rela_ndcg_1}\footnote{Note that SciFact was regenerated with our prompt, leading to slight result discrepancies from ~\citet{dai2023llms}}, respectively. \autoref{fig: intro} shows average results of neural retrieval models across all datasets. From the results, we have the following key observations:

\paratitle{Neural models consistently exhibit source bias towards LLM-generated content.}
This bias is evident across neural retrieval models, spanning both in-domain and, more significantly, out-of-distribution datasets. This trend persists in data from both pre-LLM and LLM eras. 
Remarkably, the average $\text{Relative}~\Delta$ across all neural models on the datasets surpasses $-25\%$. 
These findings further support the findings of \citet{dai2023llms} and verify the widespread source bias in neural retrieval models, regardless of the domain or task, highlighting an urgent need to address this bias.

\paratitle{Stronger neural retrieval models exhibit more severe source bias.} 
The results illustrated in \autoref{fig: intro} underscore a significant trade-off faced by neural retrieval models: advancements in ranking performance often come with an increase in source bias. 
This trend suggests that these SOTA neural models may not fully understand semantic relevance.
Instead, these models tend to leverage inherent shortcuts to enhance performance, inadvertently leading to increased bias. 
This phenomenon suggests that attempts to boost model performance could unintentionally magnify source bias issues, underlining the challenge of advancing model capabilities without leading to severe source bias.

\newtcbox{\hlprimarytab}{on line, rounded corners, box align=base, colback=lightgreen,colframe=white,size=fbox,arc=3pt, before upper=\strut, top=-2pt, bottom=-4pt, left=-2pt, right=-2pt, boxrule=0pt}

\newtcbox{\hlsecondarytab}{on line, box align=base, colback=lightred,colframe=white,size=fbox,arc=3pt, before upper=\strut, top=-2pt, bottom=-4pt, left=-2pt, right=-2pt, boxrule=0pt}

\newcommand{\dalgshifted}{\raisebox{0.5\depth}{$\downarrow$}}
\newcommand{\daugshifted}{\raisebox{0.5\depth}{$\uparrow$}}
\newcommand{\dashifted}{\raisebox{0.5\depth}{\tiny$\downarrow$}}
\newcommand{\ualgshifted}{\raisebox{0.5\depth}{$\uparrow$}}
\newcommand{\uashifted}{\raisebox{0.5\depth}{\tiny$\uparrow$}}

\newcommand{\da}[1]{{\scriptsize\hlprimarytab{\dashifted{#1}}}}
\newcommand{\ua}[1]{{\scriptsize\hlsecondarytab{\uashifted{#1}}}}

\newcommand{\uaglg}[1]{{\hlprimarytab{\ualgshifted{#1}}}}
\newcommand{\uag}[1]{{\scriptsize\hlprimarytab{\uashifted{#1}}}}

\newcommand{\dab}[1]{{\scriptsize\hlsecondarytab{\dashifted{#1}}}}
\newcommand{\dablg}[1]{{\hlsecondarytab{\dalgshifted{#1}}}}

\newcommand{\daulg}[1]{{\hlsecondarytab{\daugshifted{#1}}}}

\begin{table}[t]
\centering
\resizebox{1.0\columnwidth}{!}{
\renewcommand{\arraystretch}{1.25}
\begin{tabular}{c|ccc|ccc}
\hline\hline
Metric   & BM25 & + CE  & + monoT5 & DRAGON & + CE  & + monoT5 \\
\hline
NDCG@1   & 48.8 & 61.1  & 59.7     & 59.6 \uag{10.8}   & 61.0 \dab{0.1}  & 59.7 \dab{0.0}    \\
$\text{Relative}~\Delta$ & 22.0 & -32.9 & -29.2    &  -36.0 \dab{58.0}  &  -36.5 \dab{3.6} &  -33.7 \dab{4.5}     \\
\hline\hline
\end{tabular}
}
\caption{Re-ranking results with the top-100 retrieved hits from a first-stage BM25 or DRAGON model.}
\label{tab: res_ana_rerakner_bias_loop}
\end{table}

\paratitle{Neural re-rankers generalize better but are still biased towards LLM-generated content.}
Neural re-rankers, while achieving superior ranking performance on most datasets and showing enhanced generalization capabilities compared to neural retrievers, are not exempt from pervasive source bias. 
Specifically, re-ranking models like CE and monoT5 exhibit a significant bias towards LLM-generated content, sometimes even more pronounced than that observed in neural retrievers. This is particularly evident in datasets such as NQ and Touché-2020, further emphasizing the widespread nature of source bias in PLM-based neural retrieval models.

\paratitle{Bias in the first retrieval stage tends to propagate and even amplify during the re-ranking stage.}
This trend is particularly notable in datasets like NFCorpus and Touché-2020, where the source bias observed in the first retrieval stage persists and tends to intensify during the second-stage re-ranking. 
Moreover, as detailed in \autoref{tab: res_ana_rerakner_bias_loop}, enhancing the efficiency of the first-stage retrieval by replacing BM25 with DRAGON does not necessarily improve performance in the subsequent re-ranking phase. However, the source bias inherent in the first-stage retriever significantly impacts and may even magnify in subsequent re-ranking. This observation underscores the critical need for developing holistic approaches to mitigate bias throughout the retrieval pipeline, thereby ensuring fairness and accuracy in the whole IR system.

\subsection{Further Analysis} \label{sec:further_analysis}

\begin{figure}[t]
    \centering
    \subfigure[BERT]{
    \includegraphics[width=0.45\columnwidth]{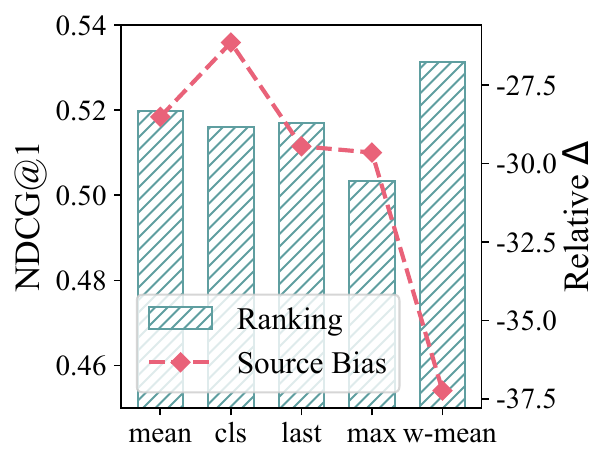}
    }
    \subfigure[RoBERTa]{
    \includegraphics[width=0.45\columnwidth]{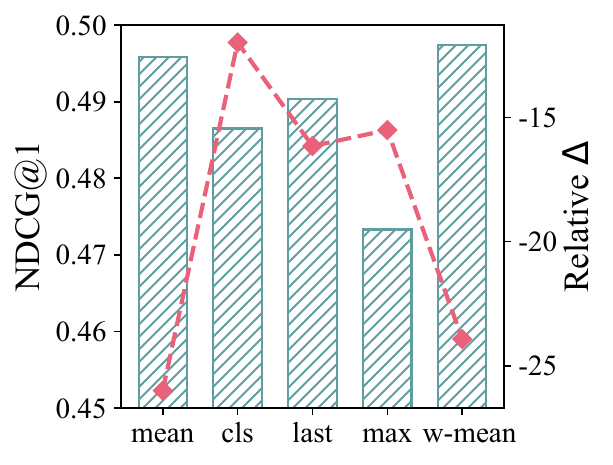}
    }
    \caption{Results of different pooling strategies. ``w-mean'' denotes weighted mean pooling. A more negative Relative $\Delta$ signifies increased source bias towards LLM-generated content.}
    \label{fig: analy_pooling}
\end{figure}

\paratitle{Impact of Different Pooling Strategies.}
Pooling strategies in PLMs are critical for aggregating information from the token embeddings for downstream semantic matching. 
We explore the ranking performance and source bias on BERT and RoBERTa w.r.t. different pooling strategies, including CLS token, max token, last token~\cite{muennighoff2022sgpt}, mean~\cite{reimers-gurevych-2019-sentence}, and weighted mean pooling~\cite{muennighoff2022sgpt}. The averaged results of all benchmarked datasets in Cocktail are shown in \autoref{fig: analy_pooling}. As we can see, the ranking performance and degree of source bias vary significantly with the specific pooling strategy.

Weighted mean pooling demonstrates the most effectiveness for ranking, which can be attributed to the nuanced semantic understanding by positionally weighting tokens, thus enhancing document relevance matching. Yet, this strategy also incurs the most severe source bias, possibly because LLM-generated texts have distinctive structural or stylistic features that become more pronounced and amplified under weighted aggregation.

Conversely, max pooling, which selects the maximum value across each dimension from the token embeddings, appears to be the least effective. This could be due to its focus on the most dominant features within the text, potentially overlooking the broader contextual nuances captured by other strategies. The dominance of specific features might not always align with the relevance signals needed for accurate document ranking, explaining the lower performance and less bias.

Other strategies, such as mean, CLS token, and last token pooling, strike a balance by capturing overall text representations while still highlighting key tokens. However, the results show that while they are capable of supporting effective document ranking, they still suffer from severe source bias with Relative $\Delta$ far below $-10\%$.

Overall, the variance in ranking performance and source bias across pooling strategies underscores the critical role of model architecture in retrieval model design. This analysis suggests that while weighted mean pooling offers enhanced ranking capabilities, it comes with a trade-off of increased source bias. Future work can explore hybrid or innovative pooling methods to balance performance with bias mitigation.

\begin{figure}[t]
\centering
\includegraphics[width=1 \columnwidth]{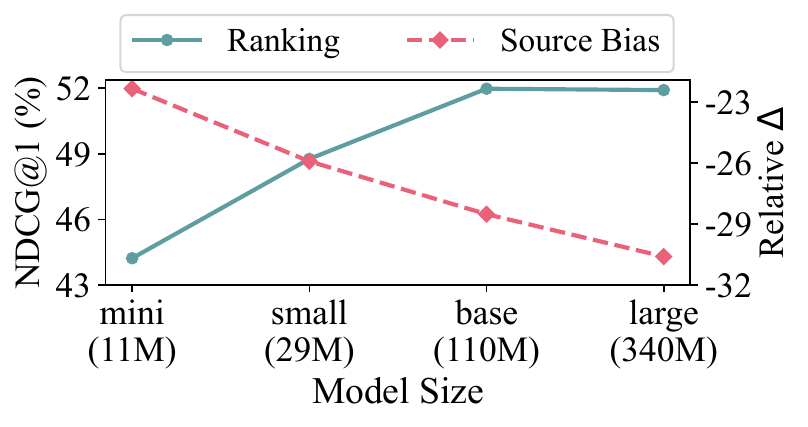}
\caption{Comparison of different model sizes. A more \textbf{negative} Relative $\Delta$ signifies increased source bias towards LLM-generated content.}
\label{fig: analy_PLM_size}
\end{figure}

\paratitle{Impact of PLM Model Size.}
Our investigation extends to the influence of model size on ranking performance and source bias, utilizing BERT models of varying sizes: mini, small, base, and large. As illustrated in \autoref{fig: analy_PLM_size}, an increase in model size leads to a notable enhancement in ranking performance. However, this improvement is paralleled by a rise in source bias (i.e., more negative).

The trend indicates that larger models, with their enhanced semantic understanding and processing capabilities, are more effective in judging document relevance, boosting their ranking performance. However, the increased capabilities may also make them more sensitive to the nuanced distinctions between human-written and LLM-generated texts, amplifying the source bias. Consequently, while the advanced performance of larger models is promising, it is coupled with a more severe bias risk, posing a significant challenge that merits more exploration in future research.

\section{Conclusion and Future Work}
This study proposes Cocktail, the first comprehensive benchmark consisting of 16 diverse datasets with mixed human-written and LLM-generated corpora. 
Alongside this, we present an evaluation tool equipped with standardized data formats and easily adaptable evaluation codes for a wide array of retrieval models.
This tool is designed to systematically evaluate ranking performance and source bias in IR systems, both in supervised and zero-shot settings, paving the way for the development of more robust next-generation IR models.

Utilizing Cocktail, we conduct an extensive evaluation of over ten state-of-the-art neural retrieval models through more than 1,000 experiments. 
Our findings reveal a clear trade-off between ranking performance and source bias in neural retrieval models, underscoring the difficulty of enhancing model performance without increasing bias towards LLM-generated content. 
This challenge emphasizes the need for a suitable balance between performance enhancement and bias mitigation in the design of future IR models.

The newly collected and annotated NQ-UTD dataset comprises queries derived from recent events, featuring all content not yet incorporated into the pre-training knowledge of most LLMs. 
This characteristic renders it a valuable resource for fairly evaluating the effectiveness of LLMs in processing new, unseen data, especially for LLM-based retrieval or question-answering systems.

\section*{Limitations}

We believe our proposed Cocktail benchmark is a foundational step toward advancing IR research in the LLM era. Nonetheless, our work still has several limitations for future research efforts. First, while our benchmarked 16 datasets encompass a broad range of IR domains and tasks, the IR field is continually evolving, with new areas of interest emerging regularly. Future updates to the Cocktail benchmark could benefit from including datasets from other IR domains, such as legal information retrieval~\cite{sansone2022legal}. This expansion would not only diversify but also enhance the benchmark’s utility across more specialized areas. Second, the scale of our NQ-UTD dataset is currently limited to 800 query-passage pairs, primarily due to the high costs of human annotation. This encompasses financial costs, the time required to develop annotation guidelines, train annotators, and perform manual audits and validations. Future initiatives could focus on expanding the NQ-UTD dataset, possibly by employing LLMs to support and streamline the annotation process~\cite{zhang-etal-2023-llmaaa}. Such an approach could facilitate broader coverage and richer labeled query-passage pairs. Third, the construction of LLM-generated corpus in our benchmarked 16 datasets was significantly influenced by the inference costs of LLM, leading us to rely exclusively on Llama2~\cite{touvron2023llama} for generating LLM-based content. However, the real-world IR landscape is shaped by a variety of LLMs. Future research might include content generated by an array of LLMs, such as OpenAI ChatGPT and Google Gemini, to mirror the diversity of LLM-generated content more accurately. Despite these limitations, we envision the Cocktail benchmark as a valuable resource for IR research in the LLM era, offering a foundation upon which next-generation IR models can be built.

\section*{Ethics Statement}

We commit to maintaining the highest ethical standards in our research, strictly adhering to ethical guidelines, applicable licenses, and laws. 
The 15 datasets utilized in this study are sourced from public repositories and have been employed within the bounds of their respective licenses and legal constraints.
For the collected NQ-UTD dataset, we followed stringent ethical protocols, ensuring the corpus collected from online news sources was carefully screened and double-checked to remove any personally identifiable or sensitive information. Participants involved in data annotation were thoroughly informed about the process and provided informed consent before their participation.

We recognize the potential risks posed by large language models (LLMs), such as generating harmful, biased, or incorrect content~\cite{pan2023risk, chen2023can}. The Llama2 model, used in this study for corpus generation, is not immune to these issues and may inadvertently produce misleading content.
We have taken measures to minimize these risks, ensuring our benchmark aligns with ethical standards.
Despite the acknowledged risks associated with LLMs, we assert that the scientific benefits and insights derived from our benchmark far outweigh potential concerns. Our resources are intended solely for scientific research, designed to foster advancements in information retrieval within this LLM era.

\section*{Acknowledgements}
This work was funded by the National Key R\&D Program of China (2023YFA1008704), the National Natural Science Foundation of China (No. 62377044, No.62276248), Beijing Key Laboratory of Big Data Management and Analysis Methods, Major Innovation \& Planning Interdisciplinary Platform for the  ``Double-First Class” Initiative, PCC@RUC, funds for building world-class universities (disciplines) of Renmin University of China, and the Youth Innovation Promotion Association CAS under Grants No.2023111.
This work was supported by the Fundamental Research Funds for the Central Universities, and the Research Funds of Renmin University of China (RUC24QSDL013). 

\bibliography{reference}
\bibliographystyle{acl_natbib}

\clearpage
\appendix
\section{Dataset Details}
In this section, we provide details about the datasets in Cocktail.
\autoref{tab:dataset_licenses} lists the dataset website links and the corresponding licenses.

\subsection{Detailed Description of Datasets} \label{app: datasets_details}
\begin{table*}[t]
    \resizebox{1.0\textwidth}{!}{
    \begin{tabular}{ l | l | l}
        \hline\hline
        \multicolumn{1}{l|}{\textbf{Dataset}} & \multicolumn{1}{c|}{\textbf{Official Website Link}} & \multicolumn{1}{c}{\textbf{License}} \\
        \hline
        MS MARCO & \url{https://microsoft.github.io/msmarco/} & MIT License \\
        DL19 & \url{https://microsoft.github.io/msmarco/TREC-Deep-Learning-2019} &  CC BY 4.0 license\\
        DL20 & \url{https://microsoft.github.io/msmarco/TREC-Deep-Learning-2020} & CC BY 4.0 license \\
        TREC-COVID & \url{https://ir.nist.gov/covidSubmit/index.html} & Dataset License Agreement \\
        NFCorpus & \url{https://www.cl.uni-heidelberg.de/statnlpgroup/nfcorpus/} & \\
        NQ & \url{https://ai.google.com/research/NaturalQuestions} & CC BY-SA 3.0 license\\
        HotpotQA & \url{https://hotpotqa.github.io} & CC BY-SA 4.0 license\\
        FiQA-2018 & \url{https://sites.google.com/view/fiqa/} & \\
        Touché-2020 & \url{https://webis.de/events/touche-20/shared-task-1.html} & CC BY 4.0 license\\
        CQADupStack & \url{http://nlp.cis.unimelb.edu.au/resources/cqadupstack/} & License 2.0 license\\
        DBPedia & \url{https://github.com/iai-group/DBpedia-Entity/} & CC BY-SA 3.0 license\\
        SCIDOCS & \url{https://allenai.org/data/scidocs} & GNU General Public License v3.0 license\\
        FEVER & \url{http://fever.ai} & CC BY-SA 3.0 license\\
        Climate-FEVER & \url{http://climatefever.ai} & \\
        SciFact & \url{https://github.com/allenai/scifact} & CC BY-NC 2.0 license\\
        NQ-UTD & \url{https://github.com/KID-22/Cocktail} & MIT License \\
        \hline\hline 
    \end{tabular}}
    \caption{Website links and licenses for the benchmarked datasets in Cocktail. (Note: Licenses for NFCorpus, FiQA-2018, and Climate-FEVER are not provided by the authors).}
    \label{tab:dataset_licenses}
\end{table*}

\newcolumntype{P}[1]{>{\centering\arraybackslash}p{#1}} 

\newlength{\ScoreColWidth}
\setlength{\ScoreColWidth}{0.06\textwidth} 

\newlength{\SourceColWidth}
\setlength{\SourceColWidth}{0.07\textwidth} 

\begin{table*}[t]
\centering
\resizebox{0.9\textwidth}{!}{
\begin{tabular}{c|cc|P{\ScoreColWidth}P{\ScoreColWidth}P{\ScoreColWidth}|P{\SourceColWidth}P{\SourceColWidth}P{\SourceColWidth}P{\SourceColWidth}P{\SourceColWidth}P{\SourceColWidth}}
\hline\hline
\multirow{2}{*}{Domain}     & \multirow{2}{*}{\# Queries} & \multirow{2}{*}{\# Passages} & \multicolumn{3}{c|}{\# Relevancy Scores}    & \multicolumn{6}{c}{\# Source}           \\
           &            &             & 0  & 1  & 2   & Google & X  & Reddit & Wikipedia & Quora & Facebook  \\
\hline
Sports     & 12         & 120         & 66                         & 17 & 37  & 60     & 22 & 10     & 19        & 9     & 0         \\
News       & 13         & 130         & 78                         & 14 & 38  & 68     & 23 & 11     & 13        & 15    & 0         \\
Scientific & 10         & 100         & 68                         & 12 & 20  & 59     & 10 & 8      & 6         & 9     & 8         \\
Technology & 12         & 120         & 92                         & 1  & 27  & 64     & 6  & 4      & 12        & 13    & 21        \\
Autos      & 8          & 80          & 50                         & 15 & 15  & 51     & 6  & 1      & 7         & 11    & 4         \\
Music      & 10         & 100         & 60                         & 18 & 22  & 67     & 6  & 8      & 5         & 14    & 0         \\
Movies     & 7          & 70          & 41                         & 8  & 21  & 53     & 2  & 3      & 9         & 3     & 0         \\
Games      & 8          & 80          & 48                         & 9  & 23  & 56     & 3  & 9      & 8         & 4     & 0         \\
\hline
All        & 80         & 800         & 503                        & 94 & 203 & 478    & 78 & 54     & 79        & 78    & 33       \\
\hline\hline
\end{tabular}
}
\caption{Statistics of our proposed NQ-UTD dataset. The column `\# Relevancy Scores' represents the count of documents categorized by their respective relevance levels.}
\label{tab: NQ_UTD_feature}
\end{table*}

\begin{table*}[t]
\centering
\resizebox{1.0\textwidth}{!}{
\begin{tabular}{l|l|p{6cm}}
\hline\hline
Domain     & Question                                                                & Reference Answer                                                             \\
\hline
Sports     & Who is the MVP of the first NBA In-Season Tournament?                   & LeBron James                                                                 \\
Sports     & Who wins 2023 FIFA Club World Cup?                                      & Manchester City F.C.                                                         \\
News       & Where was the 44th Gulf Cooperation Council (GCC) Summit held?          & Doha, Qatar                                                                  \\
News       & Which country will take over as the rotating chair of the APEC in 2024? & Peru                                                                         \\
Scientific & Which paper won NeurIPS2023 Test of Time Award?                         & Distributed Representations of Words and Phrases and their Compositionality  \\
Scientific & Which university designed AI Coscientist?                               & Carnegie Mellon University                                                   \\
Technology & Which company released the machine learning framework MLX?              & Apple                                                                        \\
Technology & What is the maximum number of cores in Intel fifth-generation Xeon CPU? & 64                                                                           \\
Autos      & When did Xiaomi announce SU7?                                           & Dec 2023                                                                     \\
Autos      & The country with the largest automobile export volume in 2023?          & China                                                                        \\
Music      & Which singer was named the 2023 Person of the Year by Time magazine?    & Taylor Swift                                                                 \\
Music      & When did Les McCann pass away?                                          & December 29, 2023                                                            \\
Moves      & Which movie was the highest-grossing film worldwide in 2023?            & Barbie                                                                       \\
Moves      & Which actress won the Best Actress award at the European Film Awards?   & Sandra Wheeler                                                               \\
Games      & Which team is the champion of the League of Legends S13 Finals?         & SKT T1                                                                       \\
Games      & Which one is the best-selling games in the US 2023?                     & Hogwarts Legacy              \\            
\hline\hline
\end{tabular}
}
\caption{Examples of queries and reference answers from different domains of our proposed NQ-UTD dataset.}
\label{tab: nq_utd_example}
\end{table*}

\subsubsection{NQ-UTD} \label{app: our_datasets}

The \textbf{N}atural \textbf{Q}uestion-\textbf{U}p \textbf{T}o \textbf{D}ate (\textbf{NQ-UTD}) dataset comprises 80 questions focusing on recent hotspot events from November 2023 to January 2024. These questions span eight domains: Sports, News, Science, Technology, Autos, Music, Movies, and Games, with approximately 10 queries per domain related to the latest events, questioning about their times, locations, or other specifics. To gather relevant passages for each query, we searched these questions on platforms like Google, X (formerly Twitter), Reddit, Wikipedia, Quora, and Facebook.
Following the practice in \citet{sun-etal-2023-chatgpt}, we also conducted searches using keywords to retrieve passages that are partially relevant but do not directly answer the questions. We also queried these questions with the latest state-of-the-art LLMs, $\texttt{gpt-4-1106-preview}$ and $\texttt{gemini-pro}$, which have their knowledge bases updated only until April 2023. The results show their 0\% accuracy in answering questions from our test set, confirming that these LLMs had no pre-knowledge about these questions. The collected 800 documents were then manually annotated for relevance by the authors and their highly educated colleagues.
Each document received a relevance score: 0 for not relevant, 1 for partially relevant (mentioning some aspects of the query but not fully answering it), and 2 for relevant (providing a complete answer to the query). To ensure consistent and high-quality annotations, each document was reviewed by five individuals, with a majority vote determining the final label.

The statistics and examples of the NQ-UTD dataset are presented in \autoref{tab: NQ_UTD_feature} and \autoref{tab: nq_utd_example}, respectively. Note that this dataset also offers a fair evaluation of the current capabilities of the advanced LLM-based IR models. As NQ-UTD contains content not previously included in LLM training data, it can serve as a valuable resource for evaluating LLMs' ability to process and answer queries on recent events.

\subsubsection{MS MARCO}
MS MARCO~\cite{nguyen2016ms}, developed by Microsoft Research, is a cornerstone dataset in the fields of natural language processing (NLP) and information retrieval (IR).
This dataset comprises over a million user-generated questions derived from Bing search logs, tailored to advance three primary NLP tasks: document ranking, passage retrieval, and question answering.
Within our Cocktail benchmark, we specifically utilize the passage retrieval subset, which contains 532K labeled query-passage pairs. A majority of the open-sourced pre-trained transformer checkpoints have been trained using this dataset. Following previous studies~\cite{thakur2021beir, lin2023train}, we employ the MS MARCO Dev set as the supervised evaluation test set.

\subsubsection{TREC}
\paratitle{DL19.}
The DL19 (TREC Deep Learning Track 2019) dataset~\cite{craswell2020overview}, is a key resource for exploring ad hoc ranking across extensive datasets from various domains. Our study focuses on the passage retrieval task utilizing the DL19 dataset, which comprises 43 queries and uses the MS MARCO passage corpus. This allows for its application in supervised evaluation settings.

\paratitle{DL20.}
The DL20 dataset~\cite{craswell2021overview} comes from the second year of the TREC Deep Learning Track, and also focuses on ad hoc ranking through human-annotated training sets across diverse domains. We also focus on the passage ranking task featured in the DL20 dataset, which includes 54 queries and also leverages the MS MARCO passage corpus. This also allows for its application in supervised evaluation settings.

\subsubsection{BEIR}
If not specified, we utilize the preprocessed version of the following open-sourced datasets as included in the BEIR benchmark~\cite{thakur2021beir} to serve as the human-written seed datasets in our Cocktail benchmark, ensuring consistency and comparability in our evaluations.

\paratitle{TREC-COVID.}
The TREC-COVID dataset is developed by the Text REtrieval Conference (TREC)~\cite{voorhees2021trec}, specifically designed to address the challenges of biomedical information retrieval in the context of the COVID-19 pandemic. 
This dataset represents a concerted effort to harness the rapidly expanding collection of scholarly articles related to COVID-19, providing a focused resource for biomedical IR.

\paratitle{NFCorpus.}
The NFCorpus dataset~\cite{boteva2016full} serves as a detailed resource tailored for ranking tasks within the biomedical field. A distinctive aspect of NFCorpus is its meticulously curated relevance links, connecting queries directly to pertinent research articles.

\paratitle{NQ.}
The NQ (Natural Questions) dataset aims to propel advancements in natural language understanding, particularly focusing on the question-answering task~\cite{kwiatkowski2019natural}. This dataset is compiled from real questions submitted by users on Google search, each annotated with answers extracted from Wikipedia articles.

\paratitle{HotpotQA.}
HotpotQA~\cite{yang2018hotpotqa} is a large-scale dataset that includes diverse questions posed by human participants. These questions necessitate a multi-step reasoning process, often requiring the extraction of answers from extensive Wikipedia texts and the identification of supporting evidence within them. HotpotQA is specifically crafted to evaluate the effectiveness of models in deciphering complex, multi-hop questions and delivering precise answers.

\paratitle{FiQA-2018.}
FiQA-2018 (Financial Opinion Mining and Question Answering-2018)~\cite{maia201818} presents a specialized question-answering challenge within the financial domain. The dataset benefits from a comprehensive corpus gathered from esteemed financial news platforms, analyst reports, and influential financial blogs.

\paratitle{Touché-2020.}
Touché-2020 dataset~\cite{bondarenko2020overview} was created with the specific aim of advancing the field of argument retrieval. Its purpose is to identify argumentative content relevant to contentious queries. It contains argumentative passages from a broad spectrum of domains, which are carefully selected and annotated.

\paratitle{CQADupStack.}
CQADupStack~\cite{hoogeveen2015cqadupstack} is a popular dataset for duplicate question retrieval, which aims to identify duplicate questions in community question
answering (cQA) forums.
This dataset includes carefully annotated and categorized questions and answers across various domains. To maintain focus without losing the essence of generalizability, our study incorporates a subset from the English domains as our selected CQADupStack dataset.

\paratitle{DBPedia.}
The DBPedia dataset~\cite{hasibi2017dbpedia} includes structured information about entities, concepts, categories, and their relationships, gathered from Wikipedia entries. 
This dataset focuses on the entity retrieval task, where queries aim to retrieve relevant entities from the English DBpedia corpus dated October 2015.

\paratitle{SCIDOCS.}
The SCIDOCS dataset~\cite{cohan2020specter} is a vast collection of scholarly articles from the Semantic Scholar database, enhanced with detailed metadata extraction and annotations. It serves as a robust dataset for the citation prediction task, where the objective is to identify papers cited by a given query paper title.

\paratitle{FEVER.}
The FEVER (Fact Extraction and VERification) dataset~\cite{thorne2018fever} comes from the fact-checking domain, offering a curated collection of human-labeled claims sourced from Wikipedia. Each claim is meticulously classified as Supported, Refuted, or NotEnoughInfo, setting the stage for a nuanced fact-checking task. 

\paratitle{Climate-FEVER.}
Climate-FEVER~\cite{diggelmann2020climate} is a specialized dataset focused on the area of climate science, comprising real-world claims accompanied by evidence sentences from Wikipedia. Similar to the FEVER dataset, the primary task involves evaluating whether the provided evidence supports, refutes, or lacks sufficient information to adjudicate each claim. 

\paratitle{SciFact.}
SciFact~\cite{wadden2020fact} is designed for the verification of scientific claims sourced from peer-reviewed literature, with each claim meticulously matched with corroborative or refuting evidence from related studies. The dataset comprises expert-crafted scientific claims alongside abstracts containing evidence, each annotated with labels indicating support or contradiction, and rationales explaining the basis of the evidence.

\subsection{Dataset Processing and Quality Control} \label{app: data_process}

Given the substantial computational cost associated with rewriting documents for the 16 datasets using Large Language Models (LLMs), we adopted sampling methods in line with practices from \citet{zhou2022dynamicretriever, sun2023learning}. Specifically, for datasets with an original corpus size smaller than 200,000 documents, we retained the entire corpus without filtering. For datasets exceeding 100,000 documents, we retain all candidate documents that appear in the labeled data (including train, valid, and test set). If the number of documents post-filtering fell below 100,000, we supplemented the corpus with a random selection of up to 100,000 additional documents from the remaining corpus to ensure a challenging dataset size.

Documents with text lengths between 10 and 2,000 words are retained for each corpus. This criterion was set because texts shorter than 10 words often consisted of empty texts or symbols, compromising data quality—this was particularly evident in the TREC-COVID dataset, which contained over 40,000 documents with empty text. 
Conversely, texts over 2,000 words were filtered out to satisfy the context length limitation of 4,096 tokens for rewriting with Llama2, impacting only 0.05\% of the data. These filters did not significantly alter the overall data distribution, with a total of less than 1\% data removed, and our evaluation results align closely with those reported in the BEIR benchmark~\cite{thakur2021beir} for the original datasets, as shown in \autoref{tab: main_mix_ndcg_1}. 
For most datasets, we generated LLM Corpus once. However, for MS MARCO, HotpotQA, DBPedia, and NQ-UTD, to construct a more challenging benchmark for evaluating source bias in IR models, we generated multiple versions and used BERT to filter and merge the corpus, aiming to increase the task difficulty.

When LLM is used for rewriting, LLM will inevitably refuse to rewrite due to safety constraints, especially for datasets like Touché-2020, which focus on retrieving contentious viewpoints. For such LLM-generated data, we keep the same content as the human-written counterparts. This approach allows for the easy removal of refused rewrites if necessary, ensuring both the integrity and the quality of the dataset processing and control measures implemented. The statistics of the 16 benchmarked datasets in Cocktails are summarized in \autoref{tab: dataset_stat}.

\subsection{Dataset Statistics and Analysis} \label{app: data_ana_statis}

\begin{table*}[t]
\centering
\resizebox{1.0\textwidth}{!}{
\begin{tabular}{cc|c|cccccccc|cc}
\hline\hline
\multirow{2}{*}{Model ($\rightarrow$)} & &  Lexical & \multicolumn{8}{c|}{Neural Retrievers}  & \multicolumn{2}{c}{Neural Re-rankers} \\

 & & BM25  & BERT  & RoBERTa & ANCE    & TAS-B      & Contriever & coCondenser & RetroMAE & DRAGON & CE     & monoT5   \\
\hline
\multirow[m]{2}{*}{MS MARCO}   & Human  & 38.6 & 49.9 & 49.8   & 51.4   & 54.2 & 55.3      & 54.1       & 55.2    & 60.2  & 58.3 & 56.8   \\
              & LLM    & 34.3 & 49.5 & 50.1   & 50.1   & 51.8 & 53.5      & 52.4       & 53.0    & 56.2  & 55.2 & 53.7   \\
\cline{2-13}             
\multirow[m]{2}{*}{DL19}       & Human  & 56.2 & 77.1 & 79.5   & 72.1   & 76.4 & 74.4      & 77.9       & 76.4    & 76.7  & 80.2 & 78.7   \\
              & LLM    & 51.2 & 79.8 & 73.6   & 68.2   & 73.3 & 70.9      & 79.5       & 76.0    & 79.1  & 81.0 & 80.6   \\
\cline{2-13}              
\multirow[m]{2}{*}{DL20}       & Human  & 54.6 & 74.7 & 75.3   & 73.5   & 75.9 & 73.2      & 78.7       & 73.5    & 81.8  & 79.6 & 76.9   \\
              & LLM    & 45.7 & 79.3 & 73.8   & 76.2   & 71.9 & 75.6      & 77.5       & 79.3    & 78.7  & 75.3 & 76.5   \\
\cline{2-13}
\multirow[m]{2}{*}{TREC-COVID}    & Human  & 62.0 & 63.0 & 61.0   & 74.0   & 75.0 & 64.0      & 73.0       & 79.0    & 75.0  & 68.0 & 83.0   \\
              & LLM    & 63.0 & 63.0 & 60.0   & 67.0   & 63.0 & 60.0      & 70.0       & 73.0    & 65.0  & 75.0 & 86.0   \\
\cline{2-13}              
\multirow[m]{2}{*}{NFCorpus}      & Human  & 43.8 & 38.4 & 32.5   & 33.4   & 40.9 & 42.9      & 43.0       & 40.3    & 42.6  & 49.2 & 48.9   \\
              & LLM    & 45.8 & 37.5 & 36.1   & 34.1   & 41.0 & 42.3      & 43.5       & 41.8    & 43.2  & 49.7 & 48.4   \\
\cline{2-13}              
\multirow[m]{2}{*}{NQ}            & Human  & 44.8 & 61.4 & 59.5   & 58.9   & 64.3 & 69.3      & 64.5       & 66.5    & 69.9  & 68.5 & 69.3   \\
              & LLM    & 43.5 & 60.5 & 57.8   & 58.1   & 64.0 & 68.1      & 63.8       & 66.3    & 68.9  & 68.0 & 68.0   \\
\cline{2-13}              
\multirow[m]{2}{*}{HotpotQA}   & Human  & 83.9 & 74.1 & 61.5   & 70.4   & 84.2 & 88.0      & 80.8       & 88.2    & 89.0  & 94.2 & 91.1   \\
              & LLM    & 80.7 & 73.7 & 62.3   & 70.8   & 83.4 & 87.3      & 81.1       & 87.2    & 88.2  & 92.6 & 89.6   \\
\cline{2-13}              
\multirow[m]{2}{*}{FiQA-2018}     & Human  & 23.3 & 23.3 & 24.5   & 27.8   & 28.7 & 31.3      & 27.3       & 30.1    & 35.5  & 33.0 & 40.0   \\
              & LLM    & 22.1 & 23.3 & 21.5   & 26.2   & 25.3 & 31.2      & 24.7       & 29.2    & 33.3  & 33.2 & 37.0   \\
\cline{2-13}              
\multirow[m]{2}{*}{Touché-2020}   & Human  & 52.0 & 46.9 & 40.8   & 49.0   & 55.1 & 48.0      & 38.8       & 44.9    & 43.9  & 51.0 & 53.1   \\
              & LLM    & 57.1 & 46.9 & 49.0   & 41.8   & 38.8 & 37.8      & 27.6       & 44.9    & 55.1  & 58.2 & 51.0   \\
\cline{2-13}              
\multirow[m]{2}{*}{CQADupStack}   & Human  & 28.1 & 24.3 & 24.3   & 29.9   & 23.1 & 33.6      & 32.9       & 30.5    & 36.0  & 33.9 & 35.7   \\
              & LLM    & 22.8 & 24.1 & 23.5   & 26.7   & 21.0 & 29.8      & 28.9       & 28.0    & 32.1  & 29.7 & 31.3   \\
\cline{2-13}              
\multirow[m]{2}{*}{DBPedia}       & Human  & 34.1 & 54.3 & 49.0   & 46.8   & 56.6 & 61.8      & 57.4       & 57.8    & 60.5  & 60.8 & 37.0   \\
              & LLM    & 33.9 & 55.3 & 49.4   & 48.5   & 55.8 & 62.1      & 57.8       & 56.5    & 59.6  & 59.0 & 33.0   \\
\cline{2-13}              
\multirow[m]{2}{*}{SCIDOCS}    & Human  & 16.0 & 13.5 & 11.3   & 14.4   & 15.8 & 16.7      & 14.2       & 16.1    & 17.2  & 18.8 & 19.6   \\
              & LLM    & 15.9 & 11.8 & 11.8   & 13.6   & 15.1 & 16.0      & 14.4       & 15.9    & 17.2  & 18.4 & 19.5   \\
\cline{2-13}              
\multirow[m]{2}{*}{FEVER}         & Human  & 65.2 & 79.0 & 74.4   & 79.5   & 82.3 & 85.4      & 72.2       & 87.2    & 86.8  & 89.8 & 25.5   \\
              & LLM    & 63.0 & 79.1 & 73.3   & 80.0   & 82.3 & 85.6      & 74.5       & 86.5    & 87.1  & 89.2 & 24.2   \\
\cline{2-13}              
\multirow[m]{2}{*}{Climate-FEVER} & Human  & 25.9 & 26.8 & 27.2   & 29.5   & 32.1 & 32.8      & 28.1       & 33.6    & 34.3  & 36.0 & 35.9   \\
              & LLM    & 24.3 & 25.9 & 26.2   & 28.0   & 32.6 & 31.3      & 27.6       & 31.0    & 33.2  & 34.4 & 33.9   \\
\cline{2-13}              
\multirow[m]{2}{*}{SciFact}       & Human  & 53.0 & 35.3 & 38.3   & 38.7   & 52.7 & 54.3      & 46.3       & 53.0    & 53.7  & 54.3 & 64.0   \\
              & LLM    & 56.3 & 35.3 & 39.7   & 39.3   & 50.7 & 53.0      & 46.7       & 51.7    & 52.7  & 53.0 & 63.7   \\
\cline{2-13}             
\multirow[m]{2}{*}{NQ-UTD}          & Human  & 71.9 & 75.6 & 63.1   & 76.3   & 81.3 & 77.5      & 73.1       & 80.0    & 76.9  & 88.1 & 89.4   \\
              & LLM    & 73.1 & 75.0 & 68.8   & 77.5   & 81.3 & 76.3      & 71.3       & 76.9    & 78.8  & 89.4 & 88.1   \\
\hline\hline
\end{tabular}
}
\caption{
Performance comparison (NDCG@1) of retrieval models on Cocktail benchmark using the sole human-written or LLM-generated corpus.}
\label{tab: main_sole_ndcg_1}
\end{table*}

\begin{table}[t]
\centering
\resizebox{0.9\columnwidth}{!}{
\begin{tabular}{ccc}
\hline\hline
Dataset       & Matching Pairs   & Random Pairs      \\
\hline
MS MARCO      & $0.9802$ ($\pm 0.0138$) & $0.6725$ ($\pm 0.0369$)  \\
DL19          & $0.9798$ ($\pm 0.0145$) & $0.6733$ ($\pm 0.0373$)  \\
DL20          & $0.9817$ ($\pm 0.0133$) & $0.6728$ ($\pm 0.0362$)  \\
TREC-COVID    & $0.9875$ ($\pm 0.0088$) & $0.7502$ ($\pm 0.0436$)  \\
NFCorpus      & $0.9856$ ($\pm 0.0096$) & $0.7490$ ($\pm 0.0373$)  \\
NQ            & $0.9847$ ($\pm 0.0150$) & $0.6996$ ($\pm 0.0348$)  \\
HotpotQA      & $0.9905$ ($\pm 0.0123$) & $0.7073$ ($\pm 0.0366$)  \\
FiQA-2018     & $0.9657$ ($\pm 0.0243$) & $0.7216$ ($\pm 0.0372$)  \\
Touché-2020   & $0.9693$ ($\pm 0.0262$) & $0.7274$ ($\pm 0.0359$)  \\
CQADupStack   & $0.9583$ ($\pm 0.0336$) & $0.7360$ ($\pm 0.0311$)  \\
DBPedia       & $0.9900$ ($\pm 0.0121$) & $0.7023$ ($\pm 0.0373$)  \\
SCIDOCS       & $0.9863$ ($\pm 0.0112$) & $0.7444$ ($\pm 0.0352$)  \\
FEVER         & $0.9866$ ($\pm 0.0110$) & $0.6983$ ($\pm 0.0363$)  \\
Climate-FEVER & $0.9873$ ($\pm 0.0107$) & $0.6992$ ($\pm 0.0366$)  \\
SciFact       & $0.9881$ ($\pm 0.0084$) & $0.7484$ ($\pm 0.0445$)  \\
NQ-UTD        & $0.9833$ ($\pm 0.0122$) & $0.7138$ ($\pm 0.0443$)  \\
\hline
Avg.          & $0.9816$ ($\pm 0.0150$)  & $0.7135$ ($\pm 0.0376$)  \\
\hline\hline
\end{tabular}
}
\caption{Comparison of cosine similarity scores between matching and random pairs for each dataset.}
\label{tab: calibration}
\end{table}

\begin{table}[t]
\centering
\resizebox{1.0\columnwidth}{!}{
\begin{tabular}{cccc}
\hline\hline
Dataset       & Human Doc      & LLM Doc        & Equal              \\
\hline
MS MARCO      & 0.0\% (0.0\%)  & 10.0\% (0.0\%) & 90\% (77.8\%)      \\
DL19          & 0.0\% (0.0\%)  & 5.0\% (0.0\%)  & 95.0\% (94.7\%)    \\
DL20          & 5.0\% (0.0\%)  & 0.0\% (0.0\%)  & 95.0\% (78.9\%)    \\
TREC-COVID    & 5.0\% (0.0\%)  & 5.0\% (0.0\%)  & 90\% (61.1\%)      \\
NFCorpus      & 0.0\% (0.0\%)  & 0.0\% (0.0\%)  & 100.0\% (75.0\%)   \\
NQ            & 0.0\% (0.0\%)  & 0.0\% (0.0\%)  & 100.0\% (100.0\%)  \\
HotpotQA      & 0.0\% (0.0\%)  & 0.0\% (0.0\%)  & 100.0\% (80.0\%)   \\
FiQA-2018     & 5.0\% (0.0\%)  & 0.0\% (0.0\%)  & 95.0\% (68.4\%)    \\
Touché-2020   & 0.0\% (0.0\%)  & 0.0\% (0.0\%)  & 100.0\% (70.0\%)   \\
CQADupStack   & 5.0\% (0.0\%)  & 5.0\% (0.0\%)  & 90\% (83.3\%)      \\
DBPedia       & 0.0\% (0.0\%)  & 0.0\% (0.0\%)  & 100.0\% (85.0\%)   \\
SCIDOCS       & 0.0\% (0.0\%)  & 0.0\% (0.0\%)  & 100.0\% (100.0\%)  \\
FEVER         & 5.0\% (0.0\%)  & 0.0\% (0.0\%)  & 95.0\% (84.2\%)    \\
Climate-FEVER & 0.0\% (0.0\%)  & 0.0\% (0.0\%)  & 100.0\% (90.0\%)   \\
SciFact       & 10.0\% (0.0\%) & 5.0\% (0.0\%)  & 85.0\% (64.7\%)    \\
NQ-UTD        & 0.0\% (0.0\%)  & 0.0\% (0.0\%)  & 100.0\% (90.0\%)   \\
\hline
Avg.          & 2.2\%          & 1.9\%          & 95.9\%             \\
\hline\hline
\end{tabular}
}
\caption{Human evaluation of semantic relevance between human-written and LLM-generated documents across all datasets in the Cocktail benchmark.}
\label{tab: human_eval_quality}
\end{table}

\begin{figure*}[t]
    \centering
    \subfigure[MS MARCO]{
    \includegraphics[width=0.24\textwidth]{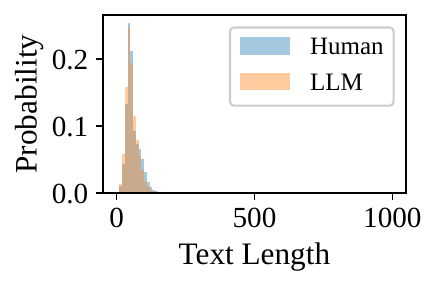}
    }
   \hspace{-0.15in}
    \subfigure[DL19]{
    \includegraphics[width=0.24\textwidth]{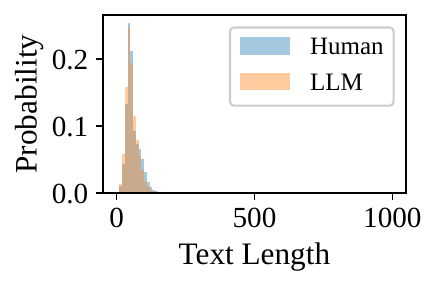}
    }
   \hspace{-0.15in}
    \subfigure[DL20]{
    \includegraphics[width=0.24\textwidth]{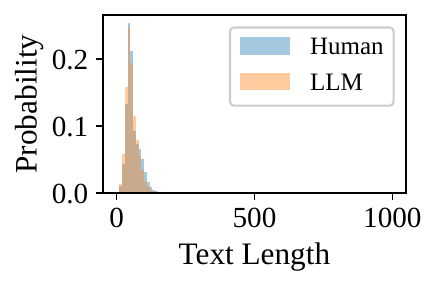}
    }
   \hspace{-0.15in}
    \subfigure[TREC-COVID]{
    \includegraphics[width=0.24\textwidth]{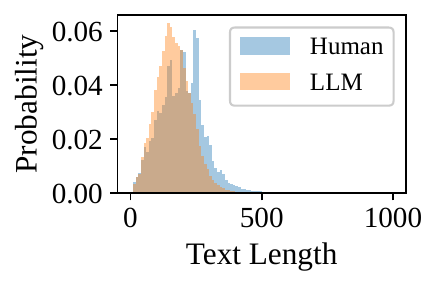}
    }
    \subfigure[NFCorpus]{
    \includegraphics[width=0.24\textwidth]{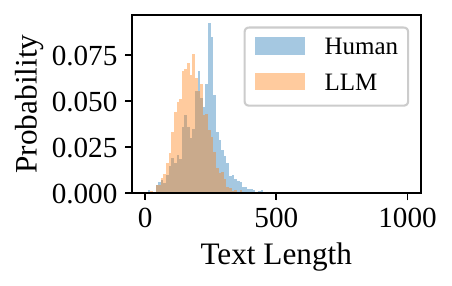}
    }
   \hspace{-0.15in}
    \subfigure[NQ]{
    \includegraphics[width=0.24\textwidth]{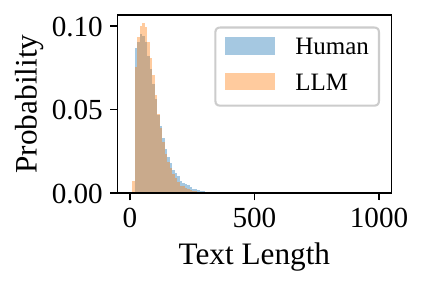}
    }
   \hspace{-0.15in}
    \subfigure[HotpotQA]{
    \includegraphics[width=0.24\textwidth]{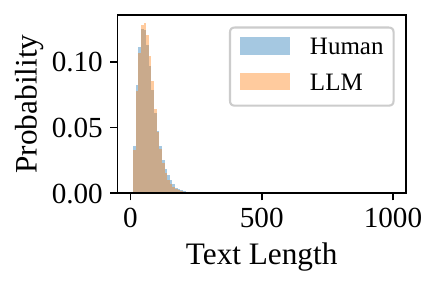}
    }
   \hspace{-0.15in}
    \subfigure[FiQA-2018]{
    \includegraphics[width=0.24\textwidth]{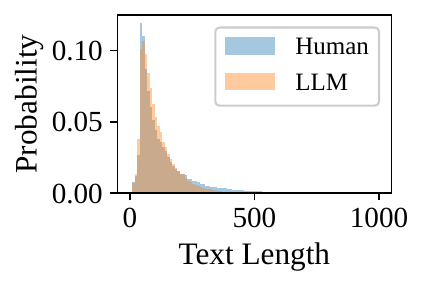}
    }
    \subfigure[Touché-2020]{
    \includegraphics[width=0.24\textwidth]{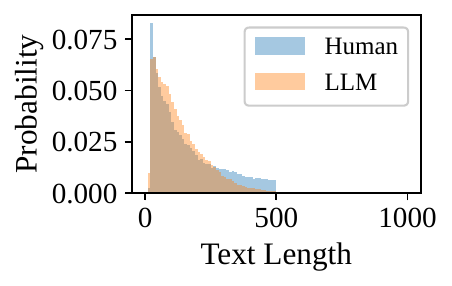}
    }
   \hspace{-0.15in}
    \subfigure[CQADupStack]{
    \includegraphics[width=0.24\textwidth]{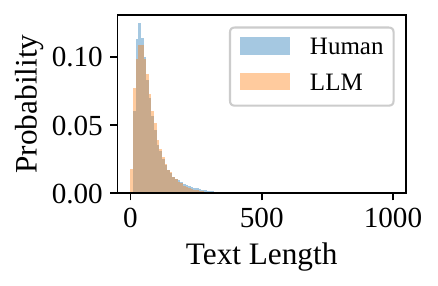}
    }
   \hspace{-0.15in}
    \subfigure[DBPedia]{
    \includegraphics[width=0.24\textwidth]{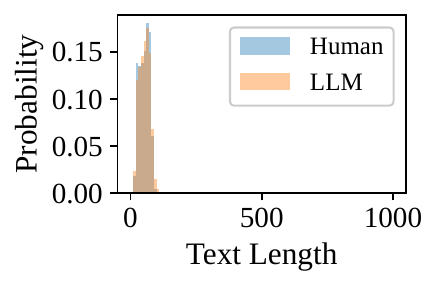}
    }
   \hspace{-0.15in}
    \subfigure[SCIDOCS]{
    \includegraphics[width=0.24\textwidth]{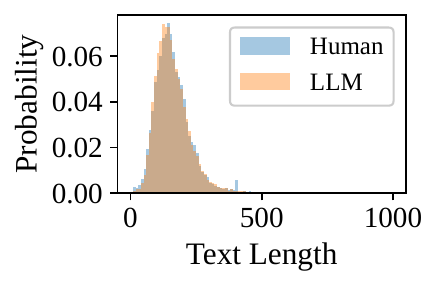}
    }
    \subfigure[FEVER]{
    \includegraphics[width=0.24\textwidth]{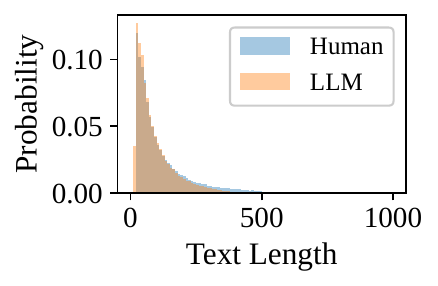}
    }
   \hspace{-0.15in}
    \subfigure[Climate-FEVER]{
    \includegraphics[width=0.24\textwidth]{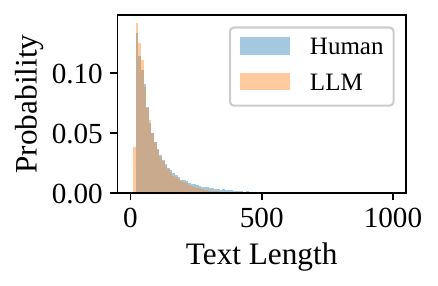}
    }
   \hspace{-0.15in}
    \subfigure[SciFact]{
    \includegraphics[width=0.24\textwidth]{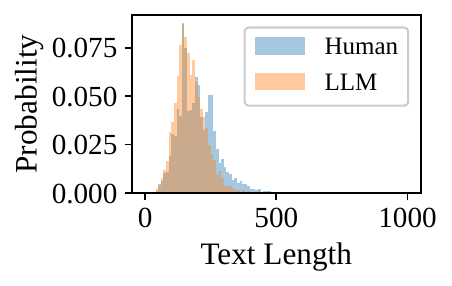}
    }
   \hspace{-0.15in}
    \subfigure[NQ-UTD]{
    \includegraphics[width=0.24\textwidth]{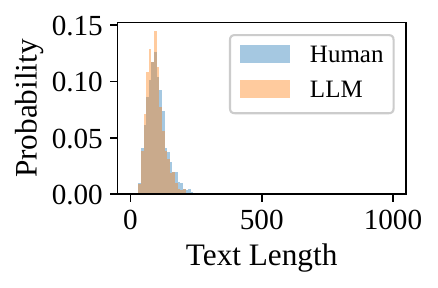}
    }
    \vspace{-0.2cm}
    \caption{Distribution of text length of corpus for each dataset in Cocktail.}
    \label{fig: stats_length}
\end{figure*}

\begin{figure*}[t]
    \centering
    \subfigure[MS MARCO]{
    \includegraphics[width=0.24\textwidth]{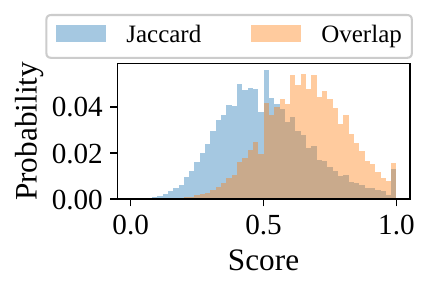}
    }
   \hspace{-0.15in}
    \subfigure[DL19]{
    \includegraphics[width=0.24\textwidth]{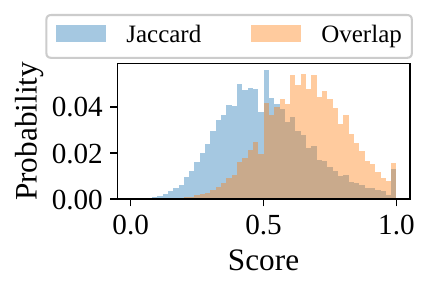}
    }
   \hspace{-0.15in}
    \subfigure[DL20]{
    \includegraphics[width=0.24\textwidth]{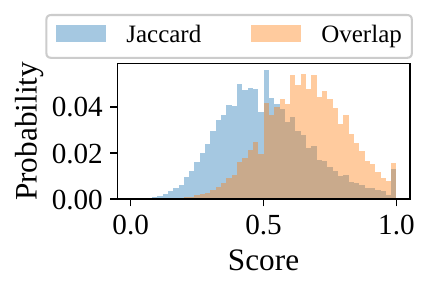}
    }
   \hspace{-0.15in}
    \subfigure[TREC-COVID]{
    \includegraphics[width=0.24\textwidth]{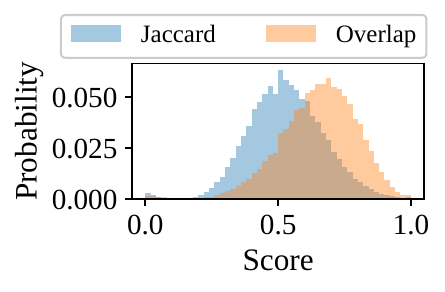}
    }
    \subfigure[NFCorpus]{
    \includegraphics[width=0.24\textwidth]{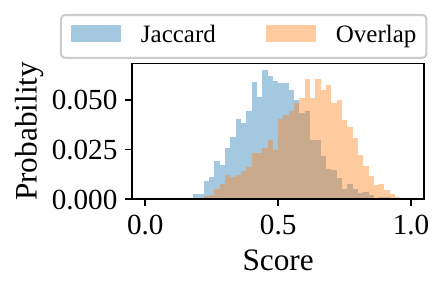}
    }
   \hspace{-0.15in}
    \subfigure[NQ]{
    \includegraphics[width=0.24\textwidth]{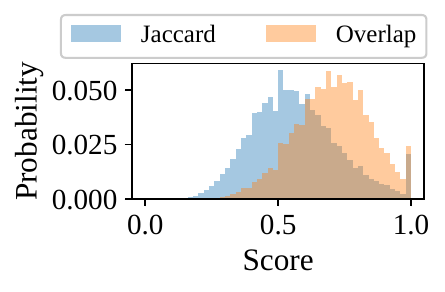}
    }
   \hspace{-0.15in}
    \subfigure[HotpotQA]{
    \includegraphics[width=0.24\textwidth]{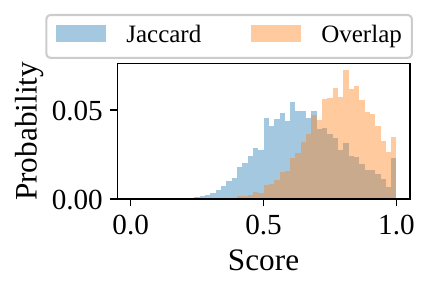}
    }
   \hspace{-0.15in}
    \subfigure[FiQA-2018]{
    \includegraphics[width=0.24\textwidth]{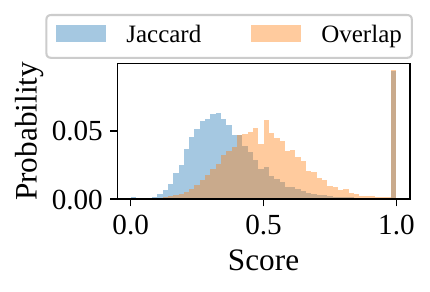}
    }
    \subfigure[Touché-2020]{
    \includegraphics[width=0.24\textwidth]{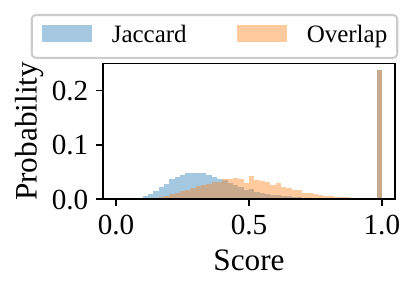}
    }
   \hspace{-0.15in}
    \subfigure[CQADupStack]{
    \includegraphics[width=0.24\textwidth]{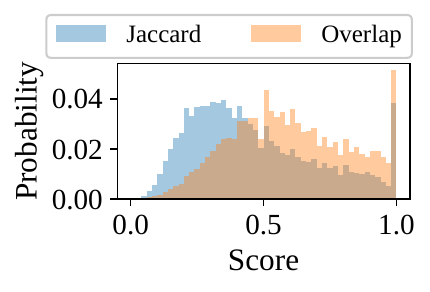}
    }
   \hspace{-0.15in}
    \subfigure[DBPedia]{
    \includegraphics[width=0.24\textwidth]{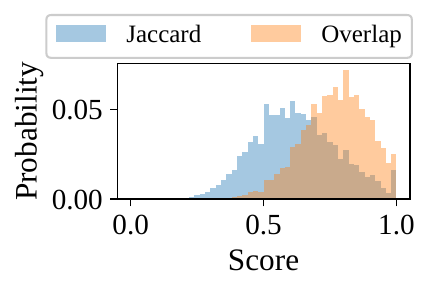}
    }
   \hspace{-0.15in}
    \subfigure[SCIDOCS]{
    \includegraphics[width=0.24\textwidth]{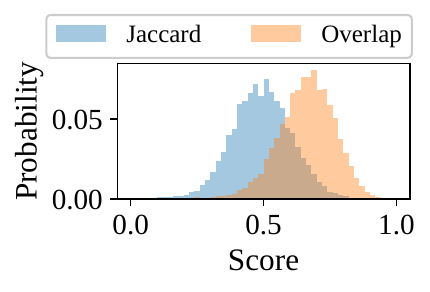}
    }
    \subfigure[FEVER]{
    \includegraphics[width=0.24\textwidth]{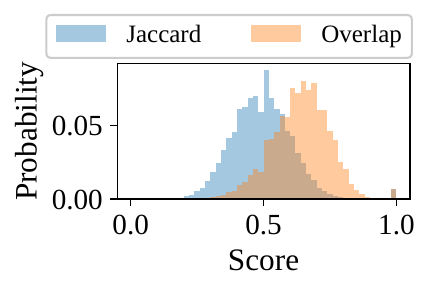}
    }
   \hspace{-0.15in}
    \subfigure[Climate-FEVER]{
    \includegraphics[width=0.24\textwidth]{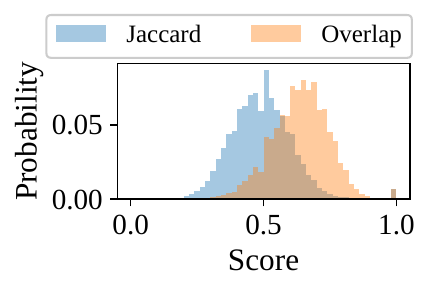}
    }
   \hspace{-0.15in}
    \subfigure[SciFact]{
    \includegraphics[width=0.24\textwidth]{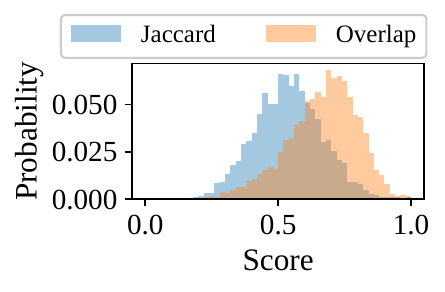}
    }
   \hspace{-0.15in}
    \subfigure[NQ-UTD]{
    \includegraphics[width=0.24\textwidth]{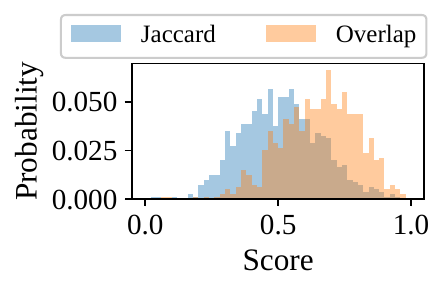}
    }
    \caption{Distribution of term-based Jaccard similarity and overlap between LLM-generated and human-written corpora for each dataset in Cocktail.}
    \label{fig: stats_jaccard_and_overlap}
\end{figure*}

\paratitle{Term-based Statistics.}
\autoref{fig: stats_length} illustrates the distribution of text lengths, revealing minimal variation between the lengths of texts. We then calculated the Jaccard similarity and overlap between each LLM-generated document $d^G$ and the corresponding human-written document $d^H$, using the following two formulas:
\begin{align}
\text{Jaccard similarity} &=\frac{|d^G \bigcap d^H|}{|d^G \bigcup d^H|},  \nonumber \\
\text{Overlap} & = \frac{|d^G \bigcap d^H|}{|d^H|}. \nonumber 
\end{align}
The results presented in \autoref{fig: stats_jaccard_and_overlap} highlight significant differences in terms despite their ostensibly similar semantic information.

\begin{figure*}[t]
    \centering
    \subfigure[MS MARCO]{
    \includegraphics[width=0.24\textwidth]{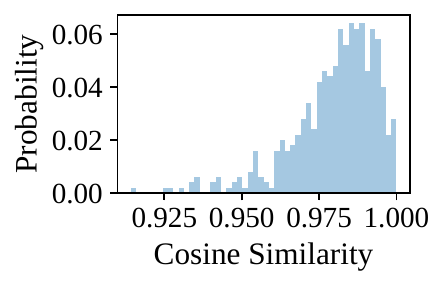}
    }
   \hspace{-0.15in}
    \subfigure[DL19]{
    \includegraphics[width=0.24\textwidth]{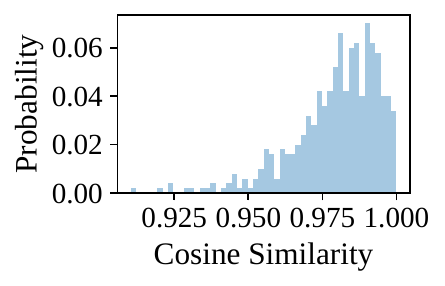}
    }
   \hspace{-0.15in}
    \subfigure[DL20]{
    \includegraphics[width=0.24\textwidth]{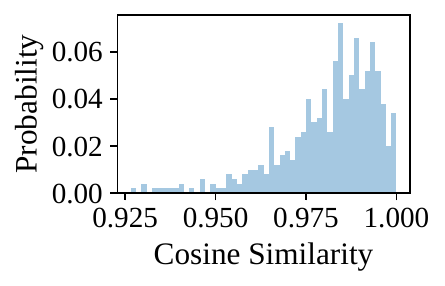}
    }
   \hspace{-0.15in}
    \subfigure[TREC-COVID]{
    \includegraphics[width=0.24\textwidth]{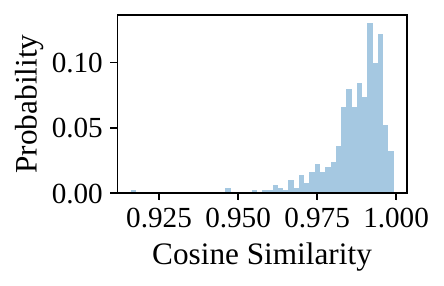}
    }
    \subfigure[NFCorpus]{
    \includegraphics[width=0.24\textwidth]{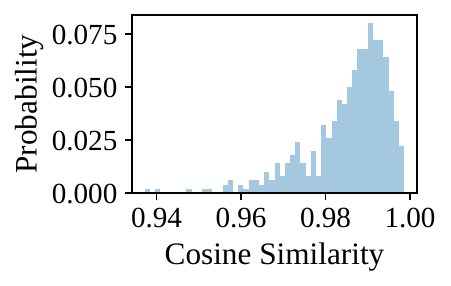}
    }
   \hspace{-0.15in}
    \subfigure[NQ]{
    \includegraphics[width=0.24\textwidth]{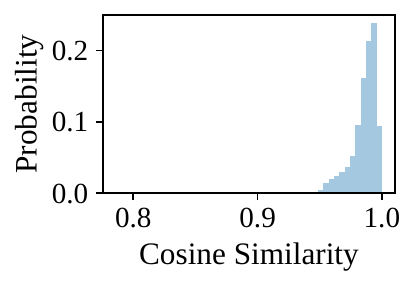}
    }
   \hspace{-0.15in}
    \subfigure[HotpotQA]{
    \includegraphics[width=0.24\textwidth]{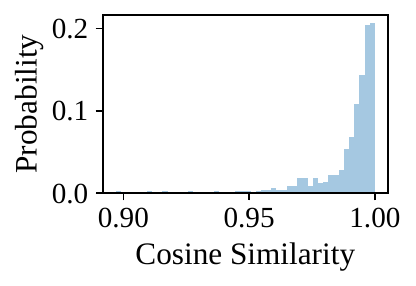}
    }
   \hspace{-0.15in}
    \subfigure[FiQA-2018]{
    \includegraphics[width=0.24\textwidth]{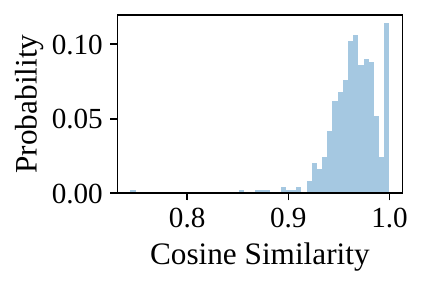}
    }
    \subfigure[Touché-2020]{
    \includegraphics[width=0.24\textwidth]{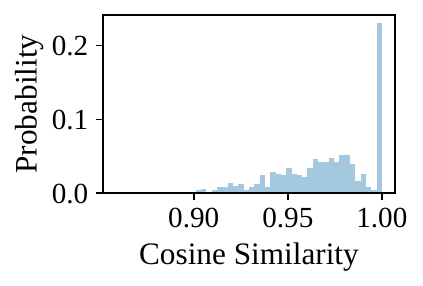}
    }
   \hspace{-0.15in}
    \subfigure[CQADupStack]{
    \includegraphics[width=0.24\textwidth]{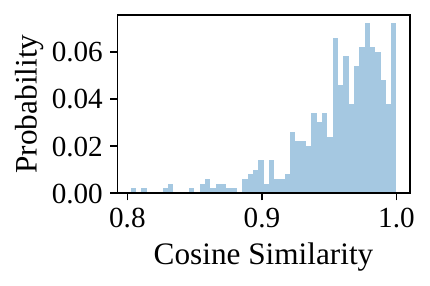}
    }
   \hspace{-0.15in}
    \subfigure[DBPedia]{
    \includegraphics[width=0.24\textwidth]{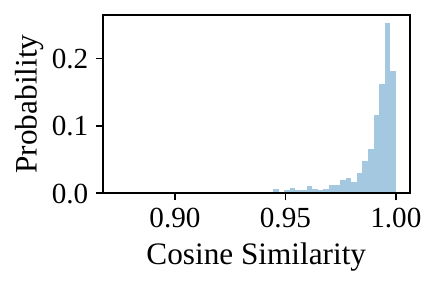}
    }
   \hspace{-0.15in}
    \subfigure[SCIDOCS]{
    \includegraphics[width=0.24\textwidth]{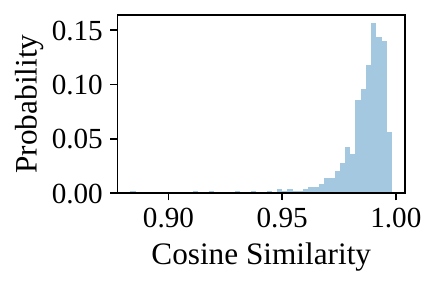}
    }
    \subfigure[FEVER]{
    \includegraphics[width=0.24\textwidth]{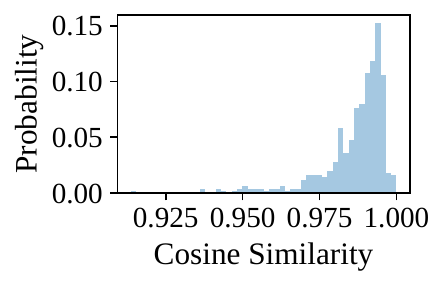}
    }
   \hspace{-0.15in}
    \subfigure[Climate-FEVER]{
    \includegraphics[width=0.24\textwidth]{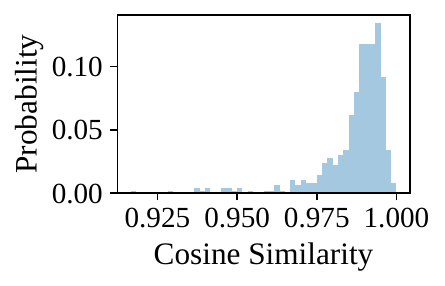}
    }
   \hspace{-0.15in}
    \subfigure[SciFact]{
    \includegraphics[width=0.24\textwidth]{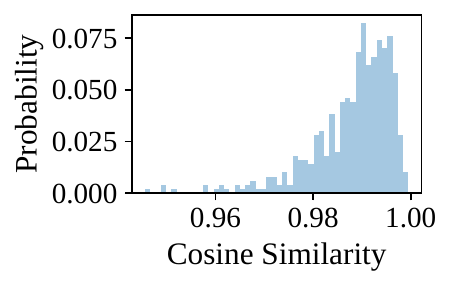}
    }
   \hspace{-0.15in}
    \subfigure[NQ-UTD]{
    \includegraphics[width=0.24\textwidth]{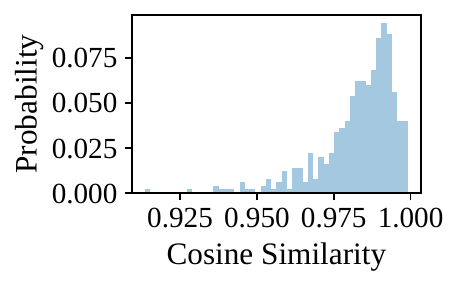}
    }
    \caption{Distribution of cosine similarity of semantic embedding between LLM-generated and human-written corpora for each dataset in Cocktail.}
    \label{fig: stats_cosine_similarity}
\end{figure*}

\paratitle{Semantic Analysis.}
Cosine similarity between LLM-generated and human-written documents, calculated using the OpenAI embedding model\footnote{\texttt{text-embedding-ada-002}}, is displayed in \autoref{fig: stats_cosine_similarity}. 
The results, predominantly above $0.9$, signify a high level of semantic preservation in the LLM-generated texts compared to the original human-written content. 
We also compare the semantics between randomly selected <human doc, LLM doc> pairs and matching pairs <human doc, LLM doc>, as shown in Table~\ref{tab: calibration}. The average similarity score for matching pairs was $0.9816$, significantly higher than the $0.7135$ average for random pairs. This significant statistical difference supports the close semantic alignment between LLM-generated and human-authored documents, further validating the quality of our constructed dataset.
Further, evaluations of various retrieval models on sole human-written or LLM-generated corpora, as detailed in \autoref{tab: main_sole_ndcg_1}, demonstrate consistent performance across all datasets between two types of corpus. These observations across all datasets reinforce confidence in the quality of our newly constructed datasets, suggesting that the LLM-generated content maintains semantics comparable to human-written texts for IR tasks.

\begin{table*}[t]
    \small
    \centering
    \ra{1.2}
    \resizebox{0.95 \textwidth}{!}{\begin{tabular}{ l | l }
        \hline
        \hline
        \multicolumn{1}{l|}{\textbf{Model Name}} & \multicolumn{1}{c}{\textbf{Publicly Available Link}} \\
        \hline
        BERT (mini) & \url{https://huggingface.co/prajjwal1/bert-mini} \\
        BERT (small) & \url{https://huggingface.co/prajjwal1/bert-small} \\
        BERT (base) & \url{https://huggingface.co/bert-base-uncased} \\
        BERT (large) & \url{https://huggingface.co/bert-large-uncased} \\
        RoBERTa & \url{https://huggingface.co/FacebookAI/roberta-base} \\
        ANCE & \url{https://huggingface.co/sentence-transformers/msmarco-roberta-base-ance-firstp} \\
        TAS-B & \url{https://huggingface.co/sentence-transformers/msmarco-distilbert-base-tas-b} \\
        Contriever & \url{https://huggingface.co/nthakur/contriever-base-msmarco} \\
        coCondenser  & \url{https://huggingface.co/sentence-transformers/msmarco-bert-co-condensor} \\
        RetroMAE & \url{https://huggingface.co/nthakur/RetroMAE_BEIR} \\
        DRAGON (Query) & \url{https://huggingface.co/nthakur/dragon-roberta-query-encoder} \\ 
        DRAGON (Doc) & \url{https://huggingface.co/nthakur/dragon-roberta-context-encoder} \\
         CE & \url{https://huggingface.co/cross-encoder/ms-marco-MiniLM-L-6-v2} \\
        monoT5 & \url{https://huggingface.co/castorini/monot5-base-msmarco-10k} \\
        \hline
        \hline
    \end{tabular}}
    \caption{Publicly available model checkpoints links used for evaluation in the Cocktail benchmark.}
    \label{tab:model_checkpoints_link}
\end{table*}

\paratitle{Quality Verification with Human Evaluation.}
To further validate the assigned relevancy label, we also conduct a human evaluation study. Due to cost constraints of human evaluation, we sample 20 <query, human doc, LLM doc> triples from each of the 16 datasets included in Cocktail. These triples were annotated by graduate students and our colleagues, who were asked to assess which document is more semantically relevant to the given query, with options being ``human doc'', ``LLM doc'', or ``equal''. At least three different annotators evaluated each triple, with the majority vote deciding the final label. We summarized the results in Table~\ref{tab: human_eval_quality}, with numbers in parentheses indicating the percentage of agreement among all three evaluators for each option.
The results show a comparable level of semantic relevance between human-written and LLM-generated texts for the given queries, ensuring the fairness of our analysis of source bias.

\subsection{Dataset Examples}

\begin{table*}[t!]
\centering
    \small
\resizebox{1.0\textwidth}{!}{
\begin{tabular}{c|p{3cm}|p{7.5cm}|p{7.5cm}}
\hline
\multicolumn{1}{c|}{\textbf{Dataset}} & \multicolumn{1}{c|}{\textbf{Query}} & \multicolumn{1}{c|}{\textbf{Relevant Human-Written Document}} & \multicolumn{1}{c}{\textbf{Relevant LLM-Generated Document}} \\

\hline

MS MARCO  &   do physicians pay for insurance from their salaries?  &  Active-duty physicians and their dependents will receive free or discounted health care coverage and dental coverage. Physicians in the Reserve and Guard can also participate in TRICARE Reserve Select, which is part of the Military's health care plan. Finally, physicians can receive up to \$400,000 in term life insurance coverage for only \$29 a month.  & Active-duty physicians and their dependents are eligible for free or discounted healthcare coverage and dental coverage. In addition, physicians in the Reserve and Guard can enroll in TRICARE Reserve Select, which is part of the Military's healthcare plan. Furthermore, physicians can receive up to \$400,000 in term life insurance coverage for just \$29 per month.  \\
\hline

DL19  &   axon terminals or synaptic knob definition  &  April 29, 2013. the terminal part of an axon from which a neural signal is rendered, via dispersion of a neurotransmitter, across a synapse to a nearby neuron. TERMINAL BUTTON: The terminal button is commonly referred to as the synaptic button, end button, button terminal, terminal bulb, and synaptic knob..  & April 29, 2013. The terminal end of an axon, where a neural signal is transmitted to a nearby neuron through the diffusion of a neurotransmitter across a synapse. TERMINAL BUTTON: Also known as the synaptic button, end button, button terminal, terminal bulb, or synaptic knob.  \\
\hline

DL20    &   average salary for dental hygienist in nebraska  &  In Nebraska, as the number of dental hygienists is growing, the salaries earned by dental hygienists are increasing. Dental hygienists earned a yearly mean salary of \$58,410 in 2006. They earned a yearly mean salary of \$64,440 in 2010.  & In Nebraska, as the number of dental hygienists is increasing, so are their salaries. According to data from 2006, dental hygienists earned a yearly mean salary of \$58,410. By 2010, this figure had risen to \$64,440.  \\
\hline

TREC-COVID  &   will SARS-CoV2 infected people develop immunity? Is cross protection possible?  &  Summary COVID-19 had a mild clinical course in patients with Agammaglobulinemia lacking B lymphocytes, whereas it developed aggressively in Common Variable Immune Deficiency. Our data offer mechanisms for possible therapeutic targets.  & COVID-19 had a mild clinical course in patients with Agammaglobulinemia, a condition characterized by a lack of B lymphocytes, while it developed aggressively in Common Variable Immune Deficiency. Our data provide insights into potential therapeutic targets.  \\
\hline

NFCorpus  &   What is Actually in Chicken Nuggets?  &  To study the origin and spread of Yersinia enterocolitica among pigs, fecal and blood samples were repeatedly taken on a fattening farm. A few piglets were found to be already infected on breeding farms. After the piglets were mixed, the infection spread through the whole unit. Eventually, all the pigs excreted the pathogen.  & To investigate the origins and spread of Yersinia enterocolitica among pigs, fecal and blood samples were repeatedly collected from a fattening farm. Initially, a few piglets on breeding farms were found to be infected. Once these piglets were mixed, the infection rapidly spread throughout the entire unit, eventually infecting all the pigs.  \\
\hline

NQ  &   how many episodes are in chicago fire season 4  &  The fourth season of Chicago Fire, an American drama television series with executive producer Dick Wolf, and producers Derek Haas, Michael Brandt, and Matt Olmstead, was ordered on February 5, 2015, by NBC,[1] and premiered on October 13, 2015 and concluded on May 17, 2016.[2] The season contained 23 episodes.[3]  & The fourth season of Chicago Fire, a US drama television series created by executive producer Dick Wolf and producers Derek Haas, Michael Brandt, and Matt Olmstead, was greenlit by NBC on February 5, 2015, and premiered on October 13, 2015, concluding on May 17, 2016. The season consisted of 23 episodes.  \\
\hline

HotpotQA  &   What government position was held by the woman who portrayed Corliss Archer in the film Kiss and Tell?  &  Kiss and Tell is a 1945 American comedy film starring then 17-year-old Shirley Temple as Corliss Archer. In the film, two teenage girls cause their respective parents much concern when they start to become interested in boys. The parents' bickering about which girl is the worse influence causes more problems than it solves.  & Kiss and Tell is a 1945 American comedy film starring then 17-year-old Shirley Temple as Corliss Archer. In the film, two teenage girls, Corliss and her friend, cause their respective parents much distress when they start to develop romantic interests in boys. The parents' arguing over which girl is the worse influence creates more problems than it solves.  \\
\hline

FiQA-2018  &   What are the ins/outs of writing equipment purchases off as business expenses in a home based business?  &  Keep this rather corny acronym in mind. Business expenses must be CORN: As other posters have already pointed out, certain expenses that are capital items (computers, furniture, etc.) must be depreciated over several years, but you have a certain amount of capital items that you can write off in the current tax year.  & Keep this straightforward acronym in mind. Business expenses must be CLEAR: As other posters have already noted, certain expenses that are capital items (computers, furniture, etc.) must be depreciated over several years, but you have a certain amount of capital items that you can write off in the current tax year.  \\
\hline

Touché-2020  &   Should teachers get tenure?  &  Why should teachers be laid off because of seniority? Many electives teachers got laid off. Even a Music Teacher got laid off even though she won a award for the best teacher in the valley! Ironic isn't it?  & Teachers should not be laid off solely based on seniority. It is unfair that many elective teachers were let go, including a talented Music Teacher who recently won an award for being the best teacher in the valley. It is ironic that she was laid off despite her achievement.  \\
\hline

CQADupStack  &   Is "a wide range of features" singular or plural?  &  I hope you can enlighten me. I get varying answers in Google and I need to find out which is the correct grammatical structure for these sentences. > The rest of the staff is/are on leave at the moment. > > The rest of my family is/are arriving late.  & I hope you can enlighten me. I've been getting conflicting answers in Google, and I need to determine the correct grammatical structure for these sentences.
The rest of the staff are on leave at the moment.
The rest of my family are arriving late.  \\
\hline

DBPedia  &   social network group selection  &  Cluster analysis or clustering is the task of grouping a set of objects in such a way that objects in the same group (called a cluster) are more similar (in some sense or another) to each other than to those in other groups (clusters).  & Clustering is the process of grouping a set of objects together based on their similarities, where objects within the same group (cluster) are more alike than those in other groups.  \\
\hline

SCIDOCS  &   DeltaCFS: Boosting Delta Sync for Cloud Storage Services by Learning from NFS  &  The state machine approach is a general method for implementing fault-tolerant services in distributed systems. This paper reviews the approach and describes protocols for two different failure models—Byzantine and fail stop. Systems reconfiguration techniques for removing faulty components and integrating repaired components are also discussed.  & The state machine approach is a versatile method for building fault-tolerant services in distributed systems. This paper examines the approach and outlines protocols for two distinct failure models—Byzantine and fail-stop. Additionally, the paper discusses techniques for reconfiguring systems to remove faulty components and integrate repaired ones.  \\
\hline

FEVER  &   Konidela Production Company was established.  &  Konidela Production Company is an Indian film production company established by actor Ram Charan , son of Chiranjeevi .  & Konidela Production Company is a renowned Indian film production house founded by the talented actor Ram Charan, the son of the legendary actor Chiranjeevi.  \\
\hline

Climate-FEVER  &   Each of the six major past ice ages began when the atmospheric carbon dioxide content was far higher than at present.  &  The Geologic temperature record are changes in Earth 's environment as determined from geologic evidence on multi-million to billion ( 109 ) year time scales . The study of past temperatures provides an important paleoenvironmental insight because it is a crucial component of the climate and oceanography of the time .  & The geologic temperature record reflects changes in Earth's environment as revealed through geologic evidence spanning millions to billions of years (109 years). The study of past temperatures offers valuable paleoenvironmental insights since it is a fundamental component of Earth's climate and oceanography during those time periods.  \\
\hline

SciFact  &   Cataract and trachoma are the primary cause of blindness in Southern Sudan.  &  Background Blindness and low vision are thought to be common in southern Sudan. However, the magnitude and geographical distribution are largely unknown. We aimed to estimate the prevalence of blindness and low vision, identify the main causes of blindness and low vision, and estimate targets for blindness prevention programs in Mankien payam (district), southern Sudan.  & Background blindness and low vision are believed to be prevalent in southern Sudan, but the scope and geographical distribution of these issues remain largely unexplored. Our study aimed to determine the prevalence of blindness and low vision, identify the primary causes of blindness and low vision, and estimate targets for blindness prevention programs in Mankien payam (district), southern Sudan.  \\
\hline

NQ-UTD  &   Who won women's doubles champion in the 2023 Badminton World Tour Finals?  &  The women's doubles final of the BWF World Tour Finals 2023 concluded at the Hangzhou Olympic Sports Center on Sunday. China's top doubles pair, Chen Qingchen and Jia Yifan, defeated South Korea's Baek Ha Na and Lee So Hee 2-0 to successfully defend their title and win \$210,000 in prize money.  & The women's doubles final of the BWF World Tour Finals 2023 came to a close at the Hangzhou Olympic Sports Center on Sunday. China's premier doubles pair, Chen Qingchen and Jia Yifan, outplayed South Korea's Baek Ha Na and Lee So Hee, winning the match 2-0 and successfully defending their title to collect a prize money of \$210,000.  \\
\hline
\end{tabular}
}    
\caption{Examples of queries, relevant human-written documents, and relevant LLM-generated documents for each dataset in our Cocktail Benchmark.}
\label{tab: dataset_examples}
\end{table*}

For a better understanding of the datasets used in our Cocktail benchmark, we offer examples for each dataset in ~\autoref{tab: dataset_examples}, showcasing a given query along with the corresponding relevant human-written document and LLM-generated document. 

\section{Model Details} \label{app: model_details}
\subsection{Detailed Description of Models}

We select the widely used approach \textbf{BM25}~\cite{robertson2009probabilistic} as the representation of \textbf{lexical retrieval models} in our benchmark.

For \textbf{neural retrieval models} that leverage pre-trained language models to acquire semantic relationships between queries and documents, we select the following representative and state-of-the-art models:

\textbf{BERT}~\cite{devlin-etal-2019-bert}, a foundational pre-trained language model, is commonly used as an encoder in dense retrieval systems. It was fine-tuned on the MSMARCO dataset using the official BEIR benchmark training script.

\textbf{RoBERTa}~\cite{liu2019roberta} builds on BERT's success with more data and refined training techniques to enhance performance. RoBERTa was fine-tuned in the same manner to BERT, using the MSMARCO dataset.

\textbf{ANCE}~\cite{xiong2020approximate} improves dense retrieval by selecting challenging negatives across the corpus and updating the Approximate Nearest Neighbor (ANN) index asynchronously with each training iteration.

\textbf{TAS-B}~\cite{hofstatter2021efficiently}, a bi-encoder model, leverages balanced margin sampling for efficient query selection and dual supervision from a cross-encoder and a ColBERT model for enhanced learning.

\textbf{Contriever}~\cite{izacard2022unsupervised} employs contrastive learning with positive samples generated through cropping and token sampling, using MoCo to incorporate negatives from a queue of previous batches for self-supervised training.

\textbf{coCondenser}~\cite{gao2022unsupervised} undertakes a two-stage process beginning with pretraining and unsupervised contrastive loss for embedding generation, followed by supervised training with the pre-trained encoder.

\textbf{RetroMAE}~\cite{xiao2022retromae} utilizes a Masked Auto-Encoder approach for pretraining, enhancing sentence reconstruction from masked inputs to refine language modeling.

\textbf{DRAGON}~\cite{lin2023train} applies progressive supervision and query augmentation for effective dense retrieval, suitable for both supervised and zero-shot settings.

Additionally, the following two widely used \textbf{neural re-ranking models} are also adopted in our evaluation framework:

\textbf{CE}~\cite{wang2020minilm} is a cross-encoder model pre-trained on MS MARCO dataset \cite{nguyen2016ms} through knowledge distillation from three teacher models: BERT-base \cite{devlin-etal-2019-bert}, BERT-large \cite{devlin-etal-2019-bert}, and ALBERT-large \cite{lan2019albert}.

\textbf{monoT5}~\cite{raffel2020exploring} is a sequence-to-sequence re-ranker based on T5~\cite{raffel2020exploring}, which is also pre-trained on the MS MARCO dataset.

\subsection{Checkpoints and Implementation Details} \label{app: ckpt_implement_details}
For the fine-tuning of BERT and RoBERTa models, we utilized the official training script\footnote{\url{https://github.com/beir-cellar/beir/blob/main/examples/retrieval/training/train_msmarco_v3.py}} in BEIR on the MSMARCO training dataset. Unless otherwise noted, mean pooling was employed as the pooling strategy, alongside cosine similarity as the scoring function. Training parameters for each model included a duration of $10$ epochs, utilizing a batch size of $75$, a learning rate set at $2e\text{-}5$, and the AdamW optimizer for efficient optimization. Our software framework utilizes PyTorch version 2.0.0 and the HuggingFace Transformers library version 4.31.0. All experiments were performed using approximately 100 hours on four NVIDIA RTX A6000 (48G) GPUs.

For the other benchmarked neural retrieval models, we use the publicly available official pre-trained checkpoints, which are listed in ~\autoref{tab:model_checkpoints_link}. For all the neural models, only the first 512 word tokens of all documents are inputted.

\section{More Experimental Results}\label{app: more_results}

\begin{table*}
\centering
\resizebox{1.0\textwidth}{!}{
\renewcommand{\arraystretch}{1.25}
\begin{tabular}{c|c|cccccccc|cc|cc}
\hline\hline
\multirow{2}{*}{Model ($\rightarrow$)} &  Lexical & \multicolumn{8}{c|}{Neural Retrievers}  & \multicolumn{2}{c|}{Neural Re-rankers} \\

     & BM25  & BERT  & RoBERTa & ANCE    & TAS-B      & Contriever & coCondenser & RetroMAE & DRAGON & CE     & monoT5  &  \\
\hline
PLM           & -     & BERT  & RoBERTa & RoBERTa & DistilBERT & BERT       &    BERT     &     BERT      &   BERT      & MiniLM & T5     &  \multicolumn{2}{c}{Average}\\
\# Paras      &  -      & 110M  & 125M    & 125M    &    66M        & 110M       &   110M   &    110M      &   110M       &    66M    &  220M      &    All     & Neural    \\
\hline
\multicolumn{14}{c}{Supervised Evaluation (In-Domain Datasets Collected in Pre-LLM Era)} \\
\hline
MS MARCO   & 39.7 & 53.5 & 53.7   & 54.6   & 57.1      & 58.3 & 57.4 & 58.1 & \textbf{61.9} & \underline{59.0}  & 58.4 & 55.6 & 57.2  \\
DL19       & 51.7 & 75.7 & 75.5   & 69.4   & 75.0      & 72.6 & 75.5 & 75.5 & \underline{76.6} & \textbf{78.8}  & 75.5 & 72.9 & 75.0  \\
DL20       & 50.7 & 75.5 & 75.5   & 72.2   & 72.9      & 71.2 & 76.5 & 77.4 & \textbf{78.0} & \underline{76.9}  & 73.0 & 72.7 & 74.9  \\
\hline
\multicolumn{14}{c}{Zero-shot Evaluation (Out-of-Domain Datasets Collected in Pre-LLM Era)} \\
\hline
TREC-COVID    & 62.4 & 66.4 & 60.0   & 71.0   & 64.0      & 63.2 & 70.9 & \underline{74.3} & 68.3 & 71.9  & \textbf{79.9} & 68.4 & 69.0  \\
NFCorpus      & \underline{44.2} & 36.3 & 32.3   & 33.0   & 39.1      & 42.1 & 41.1 & 39.3 & 41.1 & \textbf{48.2}  & 43.1 & 40.0 & 39.6  \\
NQ            & 46.2 & 62.5 & 60.0   & 60.3   & 66.3      & 70.6 & 66.0 & 67.9 & \underline{71.2} & 68.6  & \textbf{72.7} & 64.7 & 66.6  \\
HotpotQA   & 72.3 & 62.3 & 51.7   & 59.2   & 72.3      & 76.3 & 69.4 & 76.5 & 77.0 & \textbf{83.5}  & \underline{80.1} & 71.0 & 70.8  \\
FiQA-2018     & 22.2 & 21.9 & 22.2   & 26.2   & 26.1      & 29.9 & 25.3 & 28.6 & \underline{33.5} & 31.6  & \textbf{37.3} & 27.7 & 28.3  \\
Touché-2020   & 51.4 & 47.0 & 42.1   & 46.8   & 44.7      & 46.3 & 34.3 & 47.1 & \textbf{53.4} & \underline{52.8}  & 51.7 & 47.1 & 46.6  \\
CQADupStack   & 26.4 & 24.8 & 25.0   & 29.5   & 23.3      & 32.6 & 31.7 & 29.5 & \underline{34.8} & 32.6  & \textbf{35.2} & 29.6 & 29.9  \\
DBPedia       & 34.2 & 52.6 & 47.3   & 45.8   & 54.5      & \textbf{59.3} & 54.8 & 55.1 & \underline{57.7} & 57.4  & 49.6 & 51.7 & 53.4 \\
SCIDOCS    & 15.4 & 11.7 & 10.5   & 12.7   & 15.0      & 15.5 & 13.9 & 14.8 & 16.1 & \underline{17.4}  & \textbf{18.4} & 14.7 & 14.6  \\
FEVER         & 67.5 & 80.5 & 75.8   & 80.7   & 83.8      & 86.5 & 75.5 & 87.7 & \underline{87.9} & \textbf{89.1}  & 77.0 & 81.1 & 82.4  \\
Climate-FEVER & 23.4 & 25.1 & 25.2   & 25.9   & 30.3      & 29.6 & 25.4 & 29.9 & 31.0 & \textbf{32.7}  & \underline{31.8} & 28.2 & 28.7 \\
SciFact       & \underline{58.3} & 37.9 & 40.9   & 40.8   & 54.0      & 56.4 & 49.9 & 54.5 & 56.7 & 57.1  & \textbf{65.2} & 52.0 & 51.3  \\
\hline
\multicolumn{14}{c}{Zero-shot Evaluation (Out-of-Domain Datasets Collected in the LLM Era)} \\
\hline
NQ-UTD         & 70.5 & 72.4 & 66.7   & 74.3   & 78.7      & 75.4 & 69.3 & 75.5 & 76.6 & \underline{86.6}  & \textbf{87.1} & 75.7 & 76.3  \\
\hline
\multicolumn{14}{c}{Averaged Result} \\
\hline
Supervised & 47.4 & 68.2 & 68.2 & 65.4 & 68.3 & 67.4 & 69.8 & 70.3 & \textbf{72.2} & \underline{71.6} & 69.0 & 67.1 & 69.0  \\
Zero-shot  &  45.7 & 46.3 & 43.1 & 46.6 & 50.2 & 52.6 & 48.3 & 52.4 & \underline{54.3} & \textbf{56.1} & \textbf{56.1} & 50.1 & 50.6  \\
All    & 46.0 & 50.4 & 47.8   & 50.2   & 53.6      & 55.4 & 52.3 & 55.7 & 57.6 & \textbf{59.0}  & \underline{58.5} & 53.3 & 54.0 \\
\hline\hline
\end{tabular}
}
\caption{Overall ranking performance (NDCG@3) across all benchmarked datasets in Cocktail. The second-to-last column is the average result across all models, while the last column is the average for all neural retrieval models. The \textbf{best performed} result for each dataset is marked in bold, and the \underline{second best} is underlined.}
\label{tab: main_mix_ndcg_3}
\end{table*}

\begin{table*}[t]
\centering
\resizebox{1.0\textwidth}{!}{
\renewcommand{\arraystretch}{1.25}
\begin{tabular}{c|c|cccccccc|cc|cc}
\hline\hline
\multirow{2}{*}{Model ($\rightarrow$)} &  Lexical & \multicolumn{8}{c|}{Neural Retrievers}  & \multicolumn{2}{c|}{Neural Re-rankers} \\

     & BM25  & BERT  & RoBERTa & ANCE    & TAS-B      & Contriever & coCondenser & RetroMAE & DRAGON & CE     & monoT5  &  \\
\hline
PLM           & -     & BERT  & RoBERTa & RoBERTa & DistilBERT & BERT       &    BERT     &     BERT      &   BERT      & MiniLM & T5     &  \multicolumn{2}{c}{Average}\\
\# Paras      &  -      & 110M  & 125M    & 125M    &    66M        & 110M       &   110M   &    110M      &   110M       &    66M    &  220M      &    All     & Neural    \\
\hline
\multicolumn{14}{c}{Supervised Evaluation (In-Domain Datasets Collected in Pre-LLM Era)} \\
\hline
MS MARCO   & \cellcolor{pos}35.9  & \cellcolor{neg}-4.2  & \cellcolor{neg}-8.3    & \cellcolor{pos}0.5     & \cellcolor{neg}-8.5       & \cellcolor{neg}-1.7       & \cellcolor{pos}1.9         & \cellcolor{neg}-5.3     & \cellcolor{neg}-3.2   & \cellcolor{neg}-4.0   & \cellcolor{pos}0.9    & \cellcolor{pos}0.4     & \cellcolor{neg}-3.2    \\
DL19       & \cellcolor{pos}81.6  & \cellcolor{neg}-50.6 & \cellcolor{neg}-17.2   & \cellcolor{neg}-21.0   & \cellcolor{neg}-49.3      & \cellcolor{neg}-21.9      & \cellcolor{neg}-21.1       & \cellcolor{neg}-47.6    & \cellcolor{neg}-45.4  & \cellcolor{neg}-19.5  & \cellcolor{neg}-25.4  & \cellcolor{neg}-21.6   & \cellcolor{neg}-31.9   \\
DL20       & \cellcolor{pos}91.3  & \cellcolor{neg}-59.3 & \cellcolor{neg}-37.8   & \cellcolor{neg}-7.7    & \cellcolor{neg}-18.7      & \cellcolor{neg}-10.6      & \cellcolor{neg}-17.0       & \cellcolor{neg}-31.1    & \cellcolor{neg}-36.2  & \cellcolor{neg}-21.3  & \cellcolor{neg}-13.9  & \cellcolor{neg}-14.8   & \cellcolor{neg}-25.4   \\
\hline
\multicolumn{14}{c}{Zero-shot Evaluation (Out-of-Domain Datasets Collected in Pre-LLM Era)} \\
\hline
TREC-COVID    & \cellcolor{pos}25.6  & \cellcolor{neg}-51.0 & \cellcolor{neg}-40.4   & \cellcolor{neg}-33.8   & \cellcolor{neg}-73.1      & \cellcolor{neg}-62.0      & \cellcolor{neg}-68.0       & \cellcolor{neg}-25.9    & \cellcolor{neg}-17.9  & \cellcolor{neg}-66.5  & \cellcolor{neg}-46.0  & \cellcolor{neg}-41.7   & \cellcolor{neg}-48.5   \\
NFCorpus      & \cellcolor{neg}-16.2 & \cellcolor{neg}-16.8 & \cellcolor{neg}-29.7   & \cellcolor{neg}-19.1   & \cellcolor{neg}-22.7      & \cellcolor{neg}-43.2      & \cellcolor{neg}-24.9       & \cellcolor{neg}-18.9    & \cellcolor{neg}-22.9  & \cellcolor{neg}-38.4  & \cellcolor{neg}-18.5  & \cellcolor{neg}-24.7   & \cellcolor{neg}-25.5   \\
NQ            & \cellcolor{neg}-3.9  & \cellcolor{neg}-8.0  & \cellcolor{neg}-3.8    & \cellcolor{neg}-4.3    & \cellcolor{neg}-12.4      & \cellcolor{neg}-10.2      & \cellcolor{neg}-8.5        & \cellcolor{neg}-7.9     & \cellcolor{neg}-14.7  & \cellcolor{neg}-16.6  & \cellcolor{neg}-8.7   & \cellcolor{neg}-9.0    & \cellcolor{neg}-9.5    \\
HotpotQA   & \cellcolor{pos}21.0  & \cellcolor{neg}-0.5  & \cellcolor{neg}-1.8    & \cellcolor{neg}-5.7    & \cellcolor{pos}0.2        & \cellcolor{neg}-1.6       & \cellcolor{neg}-2.9        & \cellcolor{pos}1.4      & \cellcolor{neg}-1.8   & \cellcolor{pos}14.1   & \cellcolor{pos}4.9    & \cellcolor{pos}2.5     & \cellcolor{pos}0.6     \\
FiQA-2018     & \cellcolor{neg}-3.6  & \cellcolor{neg}-14.0 & \cellcolor{neg}-5.8    & \cellcolor{neg}-20.3   & \cellcolor{neg}-10.7      & \cellcolor{neg}-28.9      & \cellcolor{neg}-19.0       & \cellcolor{neg}-20.0    & \cellcolor{neg}-15.1  & \cellcolor{neg}-25.4  & \cellcolor{neg}-17.2  & \cellcolor{neg}-16.4   & \cellcolor{neg}-17.6   \\
Touché-2020   & \cellcolor{neg}-32.1 & \cellcolor{neg}-41.4 & \cellcolor{neg}-40.2   & \cellcolor{neg}-12.3   & \cellcolor{neg}-9.7       & \cellcolor{neg}-46.7      & \cellcolor{pos}6.4         & \cellcolor{neg}-18.2    & \cellcolor{neg}-53.0  & \cellcolor{neg}-69.9  & \cellcolor{neg}-53.8  & \cellcolor{neg}-33.7   & \cellcolor{neg}-33.9   \\
CQADupStack   & \cellcolor{pos}17.8  & \cellcolor{neg}-20.4 & \cellcolor{neg}-15.7   & \cellcolor{neg}-0.9    & \cellcolor{neg}-1.7       & \cellcolor{neg}-4.1       & \cellcolor{pos}1.3         & \cellcolor{neg}-20.7    & \cellcolor{neg}-4.7   & \cellcolor{neg}-7.5   & \cellcolor{pos}3.5    & \cellcolor{neg}-4.8    & \cellcolor{neg}-7.1    \\
DBPedia       & \cellcolor{pos}21.3  & \cellcolor{neg}-9.9  & \cellcolor{neg}-14.1   & \cellcolor{neg}-17.7   & \cellcolor{neg}-5.1       & \cellcolor{neg}-5.3       & \cellcolor{neg}-10.8       & \cellcolor{neg}-13.1    & \cellcolor{neg}-10.9  & \cellcolor{neg}-2.6   & \cellcolor{pos}10.9   & \cellcolor{neg}-5.2    & \cellcolor{neg}-7.9    \\
SCIDOCS    & \cellcolor{pos}1.3   & \cellcolor{neg}-1.7  & \cellcolor{neg}-21.0   & \cellcolor{neg}-1.6    & \cellcolor{neg}-1.3       & \cellcolor{neg}-6.5       & \cellcolor{neg}-5.7        & \cellcolor{neg}-10.8    & \cellcolor{neg}-18.6  & \cellcolor{neg}-12.7  & \cellcolor{neg}-23.0  & \cellcolor{neg}-9.2    & \cellcolor{neg}-10.3   \\
FEVER         & \cellcolor{neg}-5.2  & \cellcolor{pos}0.2   & \cellcolor{neg}-0.2    & \cellcolor{neg}-25.8   & \cellcolor{neg}-4.5       & \cellcolor{neg}-6.3       & \cellcolor{neg}-8.0        & \cellcolor{pos}0.6      & \cellcolor{neg}-7.0   & \cellcolor{neg}-0.7   & \cellcolor{pos}6.2    & \cellcolor{neg}-4.6    & \cellcolor{neg}-4.5    \\
Climate-FEVER & \cellcolor{pos}10.3  & \cellcolor{neg}-7.4  & \cellcolor{neg}-8.1    & \cellcolor{neg}-38.4   & \cellcolor{neg}-10.2      & \cellcolor{neg}-6.9       & \cellcolor{neg}-9.3        & \cellcolor{neg}-4.6     & \cellcolor{neg}-5.0   & \cellcolor{pos}3.2    & \cellcolor{neg}-37.9  & \cellcolor{neg}-10.4   & \cellcolor{neg}-12.5   \\
SciFact       & \cellcolor{pos}0.6   & \cellcolor{neg}-12.5 & \cellcolor{neg}-2.5    & \cellcolor{neg}-5.9    & \cellcolor{neg}-16.4      & \cellcolor{pos}2.2        & \cellcolor{neg}-10.3       & \cellcolor{neg}-6.2     & \cellcolor{neg}-7.5   & \cellcolor{pos}1.3    & \cellcolor{neg}-8.4   & \cellcolor{neg}-6.0    & \cellcolor{neg}-6.6    \\
\hline
\multicolumn{14}{c}{Zero-shot Evaluation (Out-of-Domain Datasets Collected in the LLM Era)} \\
\hline
NQ-UTD         & \cellcolor{pos}9.0   & \cellcolor{neg}-13.3 & \cellcolor{neg}-14.2   & \cellcolor{neg}-13.9   & \cellcolor{neg}-9.7       & \cellcolor{neg}-17.5      & \cellcolor{neg}-10.7       & \cellcolor{neg}-18.7    & \cellcolor{neg}-20.6  & \cellcolor{neg}-21.8  & \cellcolor{neg}-9.0   & \cellcolor{neg}-12.8   & \cellcolor{neg}-14.9   \\
\hline
\multicolumn{14}{c}{Averaged Result} \\
\hline
Supervised    & \cellcolor{pos}69.6  & \cellcolor{neg}-38.0 & \cellcolor{neg}-21.1   & \cellcolor{neg}-9.4    & \cellcolor{neg}-25.5      & \cellcolor{neg}-11.4      & \cellcolor{neg}-12.1       & \cellcolor{neg}-28.0    & \cellcolor{neg}-28.3  & \cellcolor{neg}-14.9  & \cellcolor{neg}-12.8  & \cellcolor{neg}-12.0   & \cellcolor{neg}-20.2   \\
Zero-shot     & \cellcolor{pos}3.5   & \cellcolor{neg}-15.1 & \cellcolor{neg}-15.2   & \cellcolor{neg}-15.4   & \cellcolor{neg}-13.6      & \cellcolor{neg}-18.2      & \cellcolor{neg}-13.1       & \cellcolor{neg}-12.5    & \cellcolor{neg}-15.4  & \cellcolor{neg}-18.7  & \cellcolor{neg}-15.2  & \cellcolor{neg}-13.5   & \cellcolor{neg}-15.2   \\
All      & \cellcolor{pos}15.9  & \cellcolor{neg}-19.4 & \cellcolor{neg}-16.3   & \cellcolor{neg}-14.2   & \cellcolor{neg}-15.9      & \cellcolor{neg}-16.9      & \cellcolor{neg}-12.9       & \cellcolor{neg}-15.4    & \cellcolor{neg}-17.8  & \cellcolor{neg}-18.0  & \cellcolor{neg}-14.7  & \cellcolor{neg}-13.2   & \cellcolor{neg}-16.2  \\
\hline\hline
\end{tabular}
}
\caption{Overall source bias evaluation w.r.t. $\text{Relative}~\Delta$ (NDCG@3) across all benchmarked datasets in Cocktail. The \colorbox{pos}{numbers} (i.e., $\text{Relative}~~\Delta > 0$) suggest that retrieval models generally prefer human-written content while the \colorbox{neg}{numbers} (i.e., $\text{Relative}~~\Delta \leq 0$) indicate retrieval models prefer LLM-generated content.}
\label{tab: main_rela_ndcg_3}
\end{table*}

\begin{table*}
\centering
\resizebox{1.0\textwidth}{!}{
\renewcommand{\arraystretch}{1.25}
\begin{tabular}{c|c|cccccccc|cc|cc}
\hline\hline
\multirow{2}{*}{Model ($\rightarrow$)} &  Lexical & \multicolumn{8}{c|}{Neural Retrievers}  & \multicolumn{2}{c|}{Neural Re-rankers} \\

     & BM25  & BERT  & RoBERTa & ANCE    & TAS-B      & Contriever & coCondenser & RetroMAE & DRAGON & CE     & monoT5  &  \\
\hline
PLM           & -     & BERT  & RoBERTa & RoBERTa & DistilBERT & BERT       &    BERT     &     BERT      &   BERT      & MiniLM & T5     &  \multicolumn{2}{c}{Average}\\
\# Paras      &  -      & 110M  & 125M    & 125M    &    66M        & 110M       &   110M   &    110M      &   110M       &    66M    &  220M      &    All     & Neural      \\
\hline
\multicolumn{14}{c}{Supervised Evaluation (In-Domain Datasets Collected in Pre-LLM Era)} \\
\hline
MS MARCO   & 43.5 & 58.0 & 58.1   & 59.0   & 61.7 & 63.0      & 62.0       & 62.6    & \textbf{66.6}  & 62.7 & \underline{63.1} & 60.0 & 61.7  \\
DL19       & 49.9 & \underline{75.3} & 73.4   & 69.5   & 74.3 & 71.8      & 75.0       & 73.9    & \textbf{76.7}  & \textbf{76.7} & 73.4 & 71.8 & 74.0  \\
DL20       & 48.3 & 74.9 & 73.9   & 71.6   & 73.7 & 70.7      & 75.0       & \underline{76.0}    & \textbf{77.9}  & 74.6 & 70.7 & 71.6 & 73.9  \\
\hline
\multicolumn{14}{c}{Zero-shot Evaluation (Out-of-Domain Datasets Collected in Pre-LLM Era)} \\
\hline
TREC-COVID    & 61.3 & 63.2 & 57.7   & 68.1   & 64.4 & 62.0      & 70.7       & \underline{74.4}    & 68.4  & 70.9 &\textbf{80.0} & 67.4 & 68.0  \\
NFCorpus      & \underline{41.5} & 34.3 & 30.0   & 30.1   & 37.5 & 39.7      & 38.2       & 37.3    & 39.7  & \textbf{45.4} & 41.3 & 37.7 & 37.3  \\
NQ            & 49.3 & 64.8 & 62.3   & 62.8   & 68.9 & 73.5      & 68.7       & 70.4    & \underline{73.7}  & 70.2 & \textbf{75.3} & 67.3 & 69.1  \\
HotpotQA   & 67.8 & 56.6 & 46.8   & 53.7   & 67.4 & 71.4      & 64.0       & 71.6    & 72.0  & \textbf{79.0} & \underline{76.2} & 66.0 & 65.9  \\
FiQA-2018     & 21.6 & 21.1 & 22.1   & 25.5   & 25.7 & 29.2      & 24.4       & 27.8    & \underline{32.3}  & 30.5 & \textbf{36.5} & 27.0 & 27.5 \\
Touché-2020   & \underline{50.4} & 45.6 & 40.9   & 48.7   & 42.9 & 43.7      & 32.6       & 44.4    & 50.1  & 50.2 & \textbf{50.6} & 45.5 & 45.0 \\
CQADupStack   & 27.3 & 25.4 & 25.7   & 30.1   & 24.5 & 33.5      & 32.6       & 30.2    & \underline{35.7}  & 33.1 & \textbf{36.1} & 30.4 & 30.7  \\
DBPedia       & 33.3 & 49.2 & 44.4   & 43.5   & 51.8 & \textbf{55.5}      & 52.6       & 52.6    & 54.0  & \underline{55.1} & 48.1 & 49.1 & 50.7  \\
SCIDOCS    & 14.1 & 10.7 & 9.5    & 11.4   & 13.3 & 14.2      & 12.4       & 13.6    & 14.8  & \underline{15.8} & \textbf{16.7} & 13.3 & 13.2  \\
FEVER         & 71.4 & 82.5 & 78.1   & 82.4   & 85.8 & 88.1      & 78.5       & 89.2    & \underline{89.5}  & \textbf{89.9} & 79.5 & 83.2 & 84.3  \\
Climate-FEVER & 22.1 & 23.8 & 23.9   & 24.0   & 28.6 & 28.0      & 23.7       & 27.9    & 29.0  & \textbf{31.3} & \underline{31.0} & 26.7 & 27.1  \\
SciFact       & \underline{62.0} & 41.0 & 43.0   & 42.9   & 56.4 & 59.6      & 53.0       & 57.1    & 59.9  & 61.2 & \textbf{68.0} & 54.9 & 54.2  \\
\hline
\multicolumn{14}{c}{Zero-shot Evaluation (Out-of-Domain Datasets Collected in the LLM Era)} \\
\hline
NQ-UTD         & 68.7 & 71.3 & 63.9   & 72.4   & 76.5 & 74.2      & 69.4       & 74.3    & 74.6  & \underline{83.9} & \textbf{84.9} & 74.0 & 74.5  \\
\hline
\multicolumn{14}{c}{Averaged Result} \\
\hline
Supervised    & 47.2 & 69.4 & 68.5 & 66.7 & 69.9 & 68.5 & 70.7 & 70.8 & \textbf{73.7} & \underline{71.3} & 69.1 & 67.8 & 69.8  \\
Zero-shot     & 45.4 & 45.3 & 42.2 & 45.8 & 49.5 & 51.7 & 47.8 & 51.6 & 53.4 & \underline{55.1} & \textbf{55.7} & 49.4 & 49.8  \\
All    & 45.8 & 49.9 & 47.1   & 49.7   & 53.3 & 54.9      & 52.1       & 55.2    & \underline{57.2}  & \textbf{58.2} & \textbf{58.2} & 52.9 & 53.6 \\
\hline\hline
\end{tabular}
}
\caption{Overall ranking performance (NDCG@5) across all benchmarked datasets in Cocktail. The second-to-last column is the average result across all models, while the last column is the average for all neural retrieval models. The best performance result for each dataset is marked in bold, and the second best is \underline{underlined}.}
\label{tab: main_mix_ndcg_5}
\end{table*}

\begin{table*}[t]
\centering
\resizebox{1.0\textwidth}{!}{
\renewcommand{\arraystretch}{1.25}
\begin{tabular}{c|c|cccccccc|cc|cc}
\hline\hline
\multirow{2}{*}{Model ($\rightarrow$)} &  Lexical & \multicolumn{8}{c|}{Neural Retrievers}  & \multicolumn{2}{c|}{Neural Re-rankers} \\

     & BM25  & BERT  & RoBERTa & ANCE    & TAS-B      & Contriever & coCondenser & RetroMAE & DRAGON & CE     & monoT5  &  \\
\hline
PLM           & -     & BERT  & RoBERTa & RoBERTa & DistilBERT & BERT       &    BERT     &     BERT      &   BERT      & MiniLM & T5     &  \multicolumn{2}{c}{Average}\\
\# Paras      &  -      & 110M  & 125M    & 125M    &    66M        & 110M       &   110M   &    110M      &   110M       &    66M    &  220M      &    All     & Neural      \\
\hline
\multicolumn{14}{c}{Supervised Evaluation (In-Domain Datasets Collected in Pre-LLM Era)} \\
\hline
MS MARCO   & \cellcolor{pos}30.5  & \cellcolor{neg}-4.0  & \cellcolor{neg}-6.6    & \cellcolor{pos}1.0     & \cellcolor{neg}-7.8  & \cellcolor{neg}-1.0       & \cellcolor{pos}2.2         & \cellcolor{neg}-4.7     & \cellcolor{neg}-2.6   & \cellcolor{neg}-2.5  & \cellcolor{pos}2.0    & \cellcolor{pos}0.6     & \cellcolor{neg}-2.4    \\
DL19       & \cellcolor{pos}59.5  & \cellcolor{neg}-37.9 & \cellcolor{neg}-27.7   & \cellcolor{neg}-17.0   & \cellcolor{neg}-39.3 & \cellcolor{neg}-20.7      & \cellcolor{neg}-16.5       & \cellcolor{neg}-38.6    & \cellcolor{neg}-48.1  & \cellcolor{neg}-21.7 & \cellcolor{neg}-23.0  & \cellcolor{neg}-21.0   & \cellcolor{neg}-29.0   \\
DL20       & \cellcolor{pos}74.1  & \cellcolor{neg}-31.8 & \cellcolor{neg}-34.1   & \cellcolor{neg}-5.6    & \cellcolor{neg}-13.7 & \cellcolor{neg}-10.6      & \cellcolor{neg}-14.0       & \cellcolor{neg}-16.8    & \cellcolor{neg}-27.8  & \cellcolor{neg}-28.9 & \cellcolor{neg}-14.3  & \cellcolor{neg}-11.2   & \cellcolor{neg}-19.8   \\
\hline
\multicolumn{14}{c}{Zero-shot Evaluation (Out-of-Domain Datasets Collected in Pre-LLM Era)} \\
\hline
TREC-COVID    & \cellcolor{pos}16.6  & \cellcolor{neg}-42.4 & \cellcolor{neg}-37.1   & \cellcolor{neg}-33.4   & \cellcolor{neg}-62.7 & \cellcolor{neg}-58.7      & \cellcolor{neg}-56.9       & \cellcolor{neg}-28.5    & \cellcolor{neg}-23.4  & \cellcolor{neg}-66.3 & \cellcolor{neg}-32.7  & \cellcolor{neg}-38.7   & \cellcolor{neg}-44.2   \\
NFCorpus      & \cellcolor{neg}-15.8 & \cellcolor{neg}-15.1 & \cellcolor{neg}-19.4   & \cellcolor{neg}-10.4   & \cellcolor{neg}-13.6 & \cellcolor{neg}-34.9      & \cellcolor{neg}-13.8       & \cellcolor{neg}-17.6    & \cellcolor{neg}-16.4  & \cellcolor{neg}-26.1 & \cellcolor{neg}-12.3  & \cellcolor{neg}-17.8   & \cellcolor{neg}-18.0   \\
NQ            & \cellcolor{neg}-2.3  & \cellcolor{neg}-6.7  & \cellcolor{neg}-3.2    & \cellcolor{neg}-2.2    & \cellcolor{neg}-9.7  & \cellcolor{neg}-8.8       & \cellcolor{neg}-7.0        & \cellcolor{neg}-5.8     & \cellcolor{neg}-11.3  & \cellcolor{neg}-14.8 & \cellcolor{neg}-7.5   & \cellcolor{neg}-7.2    & \cellcolor{neg}-7.7    \\
HotpotQA   & \cellcolor{pos}16.5  & \cellcolor{neg}-0.4  & \cellcolor{neg}-1.4    & \cellcolor{neg}-4.5    & \cellcolor{pos}0.8   & \cellcolor{neg}-0.7       & \cellcolor{neg}-1.4        & \cellcolor{pos}1.8      & \cellcolor{neg}-1.1   & \cellcolor{pos}10.2  & \cellcolor{pos}4.6    & \cellcolor{pos}2.2     & \cellcolor{pos}0.8     \\
FiQA-2018     & \cellcolor{neg}-5.2  & \cellcolor{neg}-14.1 & \cellcolor{neg}-5.1    & \cellcolor{neg}-12.2   & \cellcolor{neg}-9.2  & \cellcolor{neg}-18.8      & \cellcolor{neg}-14.5       & \cellcolor{neg}-18.2    & \cellcolor{neg}-12.2  & \cellcolor{neg}-17.8 & \cellcolor{neg}-14.1  & \cellcolor{neg}-12.9   & \cellcolor{neg}-13.6   \\
Touché-2020   & \cellcolor{neg}-32.2 & \cellcolor{neg}-35.3 & \cellcolor{neg}-42.9   & \cellcolor{neg}-19.8   & \cellcolor{neg}-32.0 & \cellcolor{neg}-34.1      & \cellcolor{neg}-8.5        & \cellcolor{neg}-19.1    & \cellcolor{neg}-45.2  & \cellcolor{neg}-55.1 & \cellcolor{neg}-55.0  & \cellcolor{neg}-34.5   & \cellcolor{neg}-34.7   \\
CQADupStack   & \cellcolor{pos}15.5  & \cellcolor{neg}-15.8 & \cellcolor{neg}-11.0   & \cellcolor{neg}-0.4    & \cellcolor{pos}0.5   & \cellcolor{neg}-3.8       & \cellcolor{pos}3.6         & \cellcolor{neg}-17.9    & \cellcolor{neg}-2.5   & \cellcolor{neg}-4.7  & \cellcolor{pos}3.3    & \cellcolor{neg}-3.0    & \cellcolor{neg}-4.9    \\
DBPedia       & \cellcolor{pos}18.3  & \cellcolor{neg}-11.8 & \cellcolor{neg}-13.1   & \cellcolor{neg}-17.7   & \cellcolor{neg}-4.4  & \cellcolor{neg}-2.2       & \cellcolor{neg}-11.1       & \cellcolor{neg}-9.0     & \cellcolor{neg}-9.4   & \cellcolor{neg}-3.9  & \cellcolor{pos}13.9   & \cellcolor{neg}-4.6    & \cellcolor{neg}-6.9    \\
SCIDOCS    & \cellcolor{pos}1.4   & \cellcolor{neg}-1.8  & \cellcolor{neg}-16.3   & \cellcolor{neg}-3.4    & \cellcolor{neg}-4.4  & \cellcolor{neg}-8.2       & \cellcolor{neg}-7.9        & \cellcolor{neg}-10.1    & \cellcolor{neg}-13.2  & \cellcolor{neg}-12.3 & \cellcolor{neg}-17.8  & \cellcolor{neg}-8.5    & \cellcolor{neg}-9.5    \\
FEVER         & \cellcolor{neg}-4.2  & \cellcolor{pos}0.3   & \cellcolor{pos}0.2     & \cellcolor{neg}-22.3   & \cellcolor{neg}-3.9  & \cellcolor{neg}-5.0       & \cellcolor{neg}-6.1        & \cellcolor{pos}0.4      & \cellcolor{neg}-6.1   & \cellcolor{neg}-0.8  & \cellcolor{pos}4.9    & \cellcolor{neg}-3.9    & \cellcolor{neg}-3.8    \\
Climate-FEVER & \cellcolor{pos}7.2   & \cellcolor{neg}-4.8  & \cellcolor{neg}-5.3    & \cellcolor{neg}-37.2   & \cellcolor{neg}-10.1 & \cellcolor{neg}-4.6       & \cellcolor{neg}-7.2        & \cellcolor{neg}-4.6     & \cellcolor{neg}-3.5   & \cellcolor{pos}2.3   & \cellcolor{neg}-28.3  & \cellcolor{neg}-8.7    & \cellcolor{neg}-10.3   \\
SciFact       & \cellcolor{pos}0.4   & \cellcolor{neg}-10.9 & \cellcolor{neg}-3.2    & \cellcolor{neg}-4.0    & \cellcolor{neg}-11.6 & \cellcolor{pos}1.2        & \cellcolor{neg}-8.7        & \cellcolor{neg}-5.2     & \cellcolor{neg}-7.0   & \cellcolor{neg}-0.6  & \cellcolor{neg}-6.4   & \cellcolor{neg}-5.1    & \cellcolor{neg}-5.6    \\
\hline
\multicolumn{14}{c}{Zero-shot Evaluation (Out-of-Domain Datasets Collected in the LLM Era)} \\
\hline
NQ-UTD         & \cellcolor{pos}4.4   & \cellcolor{neg}-12.2 & \cellcolor{neg}-5.0    & \cellcolor{neg}-7.7    & \cellcolor{neg}-8.1  & \cellcolor{neg}-9.0       & \cellcolor{neg}-5.7        & \cellcolor{neg}-16.4    & \cellcolor{neg}-13.0  & \cellcolor{neg}-16.3 & \cellcolor{neg}-4.1   & \cellcolor{neg}-8.5    & \cellcolor{neg}-9.7    \\
\hline
\multicolumn{14}{c}{Averaged Result} \\
\hline
Supervised    & \cellcolor{pos}54.7  & \cellcolor{neg}-24.6 & \cellcolor{neg}-22.8   & \cellcolor{neg}-7.2    & \cellcolor{neg}-20.3 & \cellcolor{neg}-10.8      & \cellcolor{neg}-9.4        & \cellcolor{neg}-20.0    & \cellcolor{neg}-26.2  & \cellcolor{neg}-17.7 & \cellcolor{neg}-11.8  & \cellcolor{neg}-10.5   & \cellcolor{neg}-17.1   \\
Zero-shot     & \cellcolor{pos}1.6   & \cellcolor{neg}-13.2 & \cellcolor{neg}-12.5   & \cellcolor{neg}-13.5   & \cellcolor{neg}-13.0 & \cellcolor{neg}-14.4      & \cellcolor{neg}-11.2       & \cellcolor{neg}-11.6    & \cellcolor{neg}-12.6  & \cellcolor{neg}-15.9 & \cellcolor{neg}-11.7  & \cellcolor{neg}-11.6   & \cellcolor{neg}-12.9   \\
All     & \cellcolor{pos}11.5  & \cellcolor{neg}-15.3 & \cellcolor{neg}-14.4   & \cellcolor{neg}-12.3   & \cellcolor{neg}-14.3 & \cellcolor{neg}-13.7      & \cellcolor{neg}-10.8       & \cellcolor{neg}-13.1    & \cellcolor{neg}-15.2  & \cellcolor{neg}-16.2 & \cellcolor{neg}-11.7  & \cellcolor{neg}-11.4   & \cellcolor{neg}-13.7  \\
\hline\hline
\end{tabular}
}
\caption{Overall source bias evaluation w.r.t. $\text{Relative}~\Delta$ (NDCG@5) across all benchmarked datasets in Cocktail. The \colorbox{pos}{numbers} (i.e., $\text{Relative}~~\Delta > 0$) suggest that retrieval models generally prefer human-written content while the \colorbox{neg}{numbers} (i.e., $\text{Relative}~~\Delta \leq 0$) indicate retrieval models prefer LLM-generated content.}
\label{tab: main_rela_ndcg_5}
\end{table*}

In addition to the findings presented through NDCG@1 in \autoref{tab: main_mix_ndcg_1} and \autoref{tab: main_rela_ndcg_1}, we further extend our analysis to include the results for ranking performance and source bias for NDCG@3 and NDCG@5. These results are detailed in \autoref{tab: main_mix_ndcg_3} and \autoref{tab: main_rela_ndcg_3} for NDCG@3, and \autoref{tab: main_mix_ndcg_5} and \autoref{tab: main_rela_ndcg_5} for NDCG@5, respectively. Consistent with our earlier observations in Section~\ref{sec: main_res}, results on these metrics also reveal a similar trend, underscoring the robustness of our findings.

\end{document}